\documentclass[11pt]{article}
\usepackage{graphicx}
\usepackage{amsmath}

\newcommand{\BABARPubYear}    {06}

\newcommand{\BABARConfNumber} {039}
\newcommand{\SLACPubNumber} {12032}

\input babarsym.tex

\def\Btopipi   {\ensuremath{\B \to \pi\pi}\xspace}
\def\Btopipiz   {\ensuremath{\Bpm \to \pipm\piz}\xspace}

\def\Btokpiz   {\ensuremath{\Bpm \to \Kpm\piz}\xspace}
\def\Btohpiz   {\ensuremath{\Bpm \to \hpm\piz}\xspace}

\def\Bztopizpiz   {\ensuremath{\Bz \to \piz\piz}\xspace}

\def\Bztopipi   {\ensuremath{\Bz \to \pipi}\xspace}
\def\Bztopippim   {\ensuremath{\Bz \to \pip\pim}\xspace}
\def\Bztokpi   {\ensuremath{\Bz \to \Kp\pim}\xspace}
\def\Bztokk   {\ensuremath{\Bz \to \Kp\Km}\xspace}
\def\Bztohh      {\ensuremath{\Bz\to h^+h^{\prime -}}\xspace}

\def\Btokzpi   {\ensuremath{\Bpm \to \Kz\pipm}\xspace}
\def\Bztokzpiz   {\ensuremath{\Bz \to \Kz\piz}\xspace}
\def\Btokzpi   {\ensuremath{\Bpm \to \Kz\pipm}\xspace}
\def\Btorhopi   {\ensuremath{\B \to \rho\pi}\xspace}

\def\Btorhopiz   {\ensuremath{\Bpm \to \rho^{\pm}\piz}\xspace}
\def\Bptorhoppiz   {\ensuremath{\B^+ \to \rho^+\piz}\xspace}
\def\Bztorhoppim   {\ensuremath{\Bz \to \rho^+\pim}\xspace}
\def\Bztorhopkm   {\ensuremath{\Bz \to \rho^+K^{-}}\xspace}

\def\Btag {\ensuremath{B_{\rm tag}}}
\def\Brec {\ensuremath{B_{\rm rec}}}
\def\Bflav {\ensuremath{B_{\rm flav}}}

\def\hpm    {\ensuremath{h^{\pm}}\xspace} 
\def\acp {\ensuremath{{\cal A}}\xspace}

\def\alphaeff {\ensuremath{\alpha_{\rm eff}}\xspace}
\def\de {\ensuremath{\Delta E}\xspace}
\def\cossph   {\ensuremath{|\cos{\theta_{\scriptscriptstyle S}}|\;}\xspace}
\def\fish    {\ensuremath{\cal F}\xspace}

\def\thetac {\ensuremath{\theta_{\rm C}}\xspace}

\def\sss{\scriptscriptstyle}
\def\barpd{{\raise.35ex\hbox
{${\sss (}$}}--{\raise.35ex\hbox{${\sss )}$}}}
\def\BorBbar{\hbox{$B^{0}$\kern-1.25em\raise1.5ex\hbox{\barpd}}}

\def\qq {\ensuremath{q\bar{q}}\xspace}
\def\spipi {\ensuremath{S_{\pi\pi}}\xspace}
\def\cpipi {\ensuremath{C_{\pi\pi}}\xspace}
\def\cpizpiz {\ensuremath{C_{\piz\piz}}\xspace}
\def\akpi {\ensuremath{\mathcal{A}_{K\pi}}\xspace}
\def\apipiz{\ensuremath{\mathcal{A}_{\pi\piz}}\xspace}
\def\akpiz{\ensuremath{\mathcal{A}_{K\piz}}\xspace}
\def\delalph{\ensuremath{\Delta \alpha_{\pi\pi}}\xspace}

\long\def\inst#1{\par\nobreak\kern 4pt\nobreak
    {\it #1}\par\vskip 10pt plus 3pt minus 3pt}

\begin{document}
{\pagestyle{empty}

\begin{flushright}
\babar-CONF-\BABARPubYear/\BABARConfNumber \\
SLAC-PUB-\SLACPubNumber \\ 
\end{flushright}

\par\vskip 2cm

\begin{center}
\Large \bf {\boldmath Measurement of \CP Asymmetries and Branching Fractions in $\B \to \pi\pi$ and $\B \to K \pi$ decays.}
\end{center}
\bigskip

\begin{center}
\large The \babar\ Collaboration\\
\mbox{ }\\
July 29, 2006
\end{center}
\vfill

\begin{center}
\large \bf Abstract
\end{center}
We present preliminary measurements of the \CP asymmetries and
branching fractions for $\B \to \pi\pi$ and $\B \to K \pi$ decays.  
 A total of 347 million \BB events collected by the \babar\ detector at
the \pep2\ asymmetric-energy $e^+e^-$ collider at SLAC are used for these
results. We find
\begin{align*}
   \spipi & =   -0.53 \pm 0.14 \pm 0.02 \\
   \cpipi & =   -0.16 \pm 0.11 \pm 0.03 \\
   {\cal A}_{K\pi} & = -0.108 \pm 0.024 \pm 0.008 \\
   \BR(\Bztopizpiz) & = ( 1.48 \pm 0.26 \pm 0.12 ) \times 10^{-6} \\
   \BR(\Btopipiz) & = ( 5.12 \pm 0.47 \pm 0.29 ) \times 10^{-6} \\
   \BR(\Btokpiz) & = ( 13.3 \pm 0.56 \pm 0.64 ) \times 10^{-6} \\
   \cpizpiz & =  -0.33 \pm 0.36 \pm 0.08   \\
   \apipiz & =  -0.019 \pm 0.088 \pm 0.014   \\
   \akpiz & =  0.016 \pm 0.041 \pm 0.012   
\end{align*}
The measured values of $\spipi$ and $\cpipi$ imply that \CP\ conservation in 
$\Bz\to\pip\pim$ decays is excluded at the $3.6~\sigma$ level.
From these results we present bounds on the CKM angle $\alpha$.

\vfill
\begin{center}

Submitted to the 33$^{\rm rd}$ International Conference on High-Energy Physics, ICHEP 06,\\
26 July---2 August 2006, Moscow, Russia.

\end{center}

\vspace{1.0cm}
\begin{center}
{\em Stanford Linear Accelerator Center, Stanford University, 
Stanford, CA 94309} \\ \vspace{0.1cm}\hrule\vspace{0.1cm}
Work supported in part by Department of Energy contract DE-AC03-76SF00515.
\end{center}

\newpage
} 

\begin{center}
\small

The \babar\ Collaboration,
\bigskip

%
{B.~Aubert,}
{R.~Barate,}
{M.~Bona,}
{D.~Boutigny,}
{F.~Couderc,}
{Y.~Karyotakis,}
{J.~P.~Lees,}
{V.~Poireau,}
{V.~Tisserand,}
{A.~Zghiche}
\inst{Laboratoire de Physique des Particules, IN2P3/CNRS et Universit\'e de Savoie,
 F-74941 Annecy-Le-Vieux, France }
% ...

\end{center}\newpage

\section{INTRODUCTION}
\label{sec:Introduction}
\CP-violating processes are incisive tests of the
Cabibbo-Kabayashi-Maskawa (CKM) model of quark mixing~\cite{Ckm}.
Measurements with small theoretical uncertainties in the CKM model
provide the most effective constraints on physics outside this model.
The CKM Unitarity Triangle angle $\alpha \equiv \arg\left[-V_{\rm
    td}^{}V_{\rm tb}^{*}/V_{\rm ud}^{}V_{\rm ub}^{*}\right]$ is
measured through the interference of $b \to u$ quark-level decays and
$\Bz \leftrightarrow \Bzb$ mixing. Multiple measurements of $\alpha$, with
different decays, further test the consistency of the CKM model.  The
time-dependent asymmetry in \Bztopippim is proportional to \stwoa in
the limit that only one amplitude contributes to this decay. However,
measurements of an unexpectedly large branching fraction for
\Bztopizpiz compared to that of \Btopipiz and
\Bztopippim~\cite{BRmeas} indicate that both $b \to u$ (tree) and $b
\to d$ (penguin) amplitudes, with different weak phases, are present.
Thus, the time-dependent asymmetry is modified to
\begin{equation}
\begin{split}\label{eq:asymmetry}
a(\Delta t) = \frac{|\bar{A}(\deltat)|^{2} - |A(\deltat)|^{2}}{|\bar{A}(\deltat)|^{2} + |A(\deltat)|^{2}} & 
        = \spipi \sin{(\deltamd\deltat)} -  \cpipi \cos{(\deltamd\deltat)} \\
\cpipi & = \frac{|A|^{2} - |\bar{A}|^{2}}{|A|^{2} + |\bar{A}|^{2}} \\
\spipi & = \sqrt{1 - \cpipi^{2}} \sin{(2\alpha - 2\delalph)},
\end{split}
\end{equation}
where \deltat is the difference between the proper decay times and
\deltamd is the $B$-meson mixing frequency.  Both the total phase
difference \delalph and \cpipi may 
differ from zero due to a penguin contribution to the decay amplitude
$A$.

The magnitude and relative phase of the penguin contribution to
the asymmetry may be unraveled with an isospin relation between the
three \Btopipi decays~\cite{Isospin}.  The amplitudes $A^{ij}$
for the $B\to \pi^i\pi^j$ decays satisfy the relation
\begin{equation}\label{eq:isospin}
A^{+0} = \frac{1}{\sqrt{2}}A^{+-} + A^{00},
\end{equation}
with a similar expression for the conjugate amplitudes.  The shape of the
corresponding isospin triangle is determined from measurements of the
branching fraction and time-integrated \CP asymmetry for each \Btopipi
decay. We define the direct \CP asymmetry as
\begin{equation}
\begin{split}\label{eq:directCP}
  \cpizpiz & = \frac{|A^{00}|^{2} - |\bar{A}^{00}|^{2}}{|A^{00}|^{2} +
    |\bar{A}^{00}|^{2}}  \\
  \apipiz & = \frac{|\bar{A}^{-0}|^{2} - |A^{+0}|^{2}}{|\bar{A}^{-0}|^{2} + |A^{+0}|^{2}}.
\end{split}
\end{equation}
From the difference in shape of these triangles for \Bz
and \Bzb, $\delalph$ may be determined up to a four-fold ambiguity.  No
\CP asymmetry is expected in the $\Delta I = 3/2$ decay \Btopipiz,
where no penguin amplitudes are present.

The phenomenology of the \Btopipi system has been studied in a variety
of theoretical frameworks and models~\cite{Models}.  Predictions for
the relative size and phase of the penguin contribution vary
considerably, so more precise measurements will help to
distinguish among different theoretical approaches and add to our
understanding of hadronic \B decays.

In addition to the unexpected pattern of decay rates in the $B\to\pi\pi$ 
system mentioned above, the measured rates and direct \CP-violating asymmetries 
in $B\to K\pi$ decays~\cite{ref:BaBarKpiData,ref:BelleKpiData,ref:CLEOdata} 
reveal puzzling features that could indicate significant contributions from 
electroweak (EW) penguins~\cite{ref:KpiPuzzle1,ref:KpiPuzzle2}.  
Various methods have been proposed to isolate the Standard Model contribution 
to this process in order to test for signs of new physics.  
Sum rules derived from $U$-spin symmetry relate the rates and asymmetries for 
the decays $\Bz$ or $\Bp$ to $\Kp\pim$, $\Kp\piz$, $\Kz\piz$, and $\Kz\pip$~\cite{ref:SumRule}, 
while $SU(3)$ symmetry can be used to make predictions for the $K\pi$ system
based on hadronic parameters extracted from the $\pi\pi$
system~\cite{ref:KpiPuzzle1}.

\section{THE \babar\ DETECTOR AND DATASET}
\label{sec:babar}

The data used in this analysis were collected in 1999--2006 with the
\babar\ detector at the \pep2\ asymmetric-energy \B-meson factory at the
Stanford Linear Accelerator Center. A total of 347 million \BB pairs
were used.  The preliminary results presented here supersede
the results in three prior publications~\cite{priorResults}. Roughly 120
million more \BB decays have been added, and a number of improvements
have been introduced to the data analysis, effectively increasing the
acceptance for the modes containing a neutral pion.

The \babar\ detector is described in detail
elsewhere~\cite{babar}.  Charged-particle (track) momenta are
measured with a 5-layer double-sided silicon vertex tracker (SVT) and a
40-layer drift chamber (DCH) inside a 1.5-T superconducting solenoidal
magnet.  Neutral-cluster (photon) positions and energies are measured
with an electromagnetic calorimeter (EMC) consisting of 6580 CsI(Tl)
crystals.  The photon energy resolution is $\sigma_{E}/E = \left\{2.3
  / E(\gev)^{1/4} \oplus 1.9 \right\} \%$, and the angular resolution
from the interaction point is $\sigma_{\theta} = 3.9^{\rm
  o}/\sqrt{E(\gev)}$.  Charged hadrons are identified with a detector
of internally reflected Cherenkov light (DIRC) and ionization
measurements in the tracking detectors.  The average $K$--$\pi$
separation in the DIRC varies from $12\sigma$ at a laboratory momentum
of $1.5~\gevc$ to $2\sigma$ at $4.5~\gevc$.

\section{ANALYSIS METHOD}
\label{sec:Analysis}

Many elements of the \Btopipi measurements are common to the three
groups of 
decay modes $\Bztohh (h= \pi \;{\rm or} \; K)$, \Bztopizpiz, and \Btohpiz.  
\B candidates (\Brec) are
formed by combining two particles, either tracks or \piz candidates.

\subsection{Track and Cluster Selection}
\label{sec:TrackSelection}

For the \Btohpiz and the \Bztohh samples, we require that each track
have an associated Cherenkov angle (\thetac) measured with more than
five signal photons detected in the DIRC, where the value of
\thetac must agree with either the pion or kaon particle hypothesis
to within $4\,\sigma$.  The last requirement efficiently removes events
containing high-momentum protons.  Electrons are removed based on
energy-loss measurements in the SVT and DCH, and on a comparison of
the track momentum and the associated energy deposited in the EMC.

The \piz candidates are formed from two EMC clusters, one EMC cluster
containing two nearby photons (merged \piz), or one EMC cluster and
two tracks from a photon conversion to an \epem pair inside the detector. Previous
\babar\ results for \Bztopizpiz and \Btohpiz only included \piz from
two EMC clusters; the addition of merged \piz and converted photons
increases the \piz efficiency by 10\%.  Clusters are required to have
a transverse energy deposit consistent with a photon, and to have an
energy $E_{\gamma} > 0.03~\gev$. To reduce the background from
random photon combinations, the angle $\theta_{\gamma}$ between the
photon momentum vector in the \piz rest frame and the \piz momentum
vector in the laboratory frame is required to satisfy
$|\cos{\theta_{\gamma}}| < 0.95$. The \piz\ candidates are fitted
kinematically with their mass constrained to the nominal \piz mass.

Photon conversions are selected from pairs of oppositely
charged tracks with invariant mass less than $30~\mevcc$ and whose
momentum vector points to the beamspot.  The conversion point
is required to lie inside the detector material.  Photons from
conversions are combined with photons from single EMC clusters to form
\piz candidates. 

Single EMC clusters containing two photons are selected
with the transverse second moment, $S = \sum_{i} E_{i} \times (\Delta
\alpha_i)^{2}/ E$, where $E_{i}$ is energy in each CsI(Tl) crystal and
$\Delta \alpha_{i}$ is the angle between the cluster centroid and
the crystal.  The second moment is used to distinguish merged \piz
candidates from both single photons and neutral hadrons.  

\subsection{Event Selection}
\label{sec:EventSelection}

Two kinematic variables are used to separate \B decays from the large
$\epem \to \qq \;(q=u,d,s,c)$ background: the
beam-energy--substituted mass $\mes = \sqrt{ (s/2 + {\bf
    p}_{i}\cdot{\bf p}_{B})^{2}/E_{i}^{2}- {\bf p}^{2}_{B}}$, where
$\sqrt{s}$ is the total \epem center-of-mass (CM) energy, $(E_{i},{\bf
  p}_{i})$ is the four-momentum of the initial \epem system and ${\bf
  p}_{B}$ is the \B-candidate momentum, both measured in the
laboratory frame, and \de $ = E_{B} - \sqrt{s}/2$, where $E_{B}$ is
the \B-candidate energy in the CM frame.

Two additional quantities take advantage of the event topology to
further separate \B decays from \qq background.  The cosine of the
angle $\theta_{\scriptscriptstyle S}$ between the sphericity axes of
the \B candidate's decay products and that of the remaining tracks and
clusters in the event, in the CM frame, is peaked at $\pm1.0$ for jet-like
\qq events, but has a flat distribution for \B decays.  We require
$\cossph < 0.7 \,(0.8)$ for \Bztopizpiz (\Btohpiz and \Bztohh).  For
just the $\Bztohh$ sample, we further require that the second
Fox--Wolfram moment satisfy $R_2 <0.7$ to remove a small remaining
background from $\tau$-pair events.  To improve the discrimination
against \qq events, a Fisher discriminant \fish is formed from the
sums $\sum_i p_i$ and $\sum_i p_i \cos^2{\theta_i}$, where $p_{i}$ is
the momentum and $\theta_{i}$ is the angle with respect to the thrust
axis of the \B candidate, both in the CM frame, of all tracks and
clusters not used to reconstruct the \B meson.

The number of \B decays and the corresponding \CP asymmetries are
determined in extended unbinned maximum likelihood (ML) fits to the variables
\mes, \de, and \fish, plus additional information as described below.
The likelihood is given by the expression
\begin{equation}
{\cal L} = \exp{\left(-\sum_{i}^{M} n_i \right)}
\prod_{j}^{N} \left[\sum_{i}^{M} n_i {\cal P}_{i}(\vec{x}_j;\vec{\alpha}_i)\right],
\end{equation}
where the product is over the number of events $N$, the sums are over
the event categories $M$, $n_{i}$ is the coefficient for each
category as described below, and the probability density function (PDF) $\cal P$
describes the distribution of the variables $\vec{x}$ in terms of
parameters $\vec{\alpha}$.

\subsection{\boldmath \Bztopippim and \Bztokpi}
\label{sec:hhMethod}

The time-dependent \CP asymmetry measurement in \Bztopippim also uses
\B-flavor, decay-time, and particle-identification
information to separate $\Bztopippim$ and $\Bztokpi$ decays and measure
their \CP asymmetries.  

The variables \mes and \de are calculated assuming that both tracks are
charged pions.  $\Btopipi$ events are described by a Gaussian
distribution for both \mes and \de, with resolutions of $2.5\mevcc$
and $28~\mev$, respectively.  For each kaon in the final state, the
$\de$ peak position is shifted from zero by an amount that depends on the
kaon momentum, with an average shift of $-45\mev$.  We require $5.20 <
\mes < 5.29\gevcc$ and $\left|\de\right|<0.150\gev$.  The large region
below the signal in $\mes$ effectively determines the background
shape parameters, while the wide range in $\de$ allows us to separate
$B$ decays to all four final states in the same fit.

The Cherenkov angle \thetac measured by the DIRC is used to further
separate charged kaons and pions.  The difference between the measured
and expected values of \thetac, divided by its uncertainty, is modeled
by a sum of two Gaussian distributions.  The \thetac PDF is parametrized
separately for $\Kp$, $\Km$, $\pip$, and $\pim$ tracks as a function
of momentum and polar angle.

We use a multivariate technique~\cite{BaBarsin2beta} to determine the
flavor of the other \B meson ($\Btag$).  Separate neural networks are
trained to identify primary leptons, kaons, soft pions from $D^*$
decays, and high-momentum charged particles from \B\ decays.  Events
are assigned to one of seven mutually exclusive tagging categories
(including untagged events) based on the estimated average mistag
probability and the source of the tagging information
(Table~\ref{tab:tagging}).  The quality of tagging is expressed in
terms of the effective efficiency $Q = \sum_k \epsilon_k(1-2w_k)^2$,
where $\epsilon_k$ and $w_k$ are the efficiencies and mistag
probabilities, respectively, for events tagged in category $k$. The
difference in mistag probabilities is given by $\Delta w = w_{\Bz} -
w_{\Bzb}$.  Table~\ref{tab:tagging} summarizes the tagging performance
measured in a data sample of fully reconstructed neutral $B$ decays to
$D^{(*)-}(\pip,\, \rho^+,\, a_1^+)$ (\Bflav).

The time difference $\deltat = \Delta z/\beta\gamma c$ is obtained
from the known boost of the $\epem$ system ($\beta\gamma = 0.56$) and
the measured distance $\Delta z$ along the beam ($z$) axis between the
\Brec\ and \Btag\ decay vertices.  We require
$\left|\deltat\right|<20\ps$ and $\sigma_{\deltat} < 2.5\ps$, where
$\sigma_{\deltat}$ is the error on $\deltat$ determined separately for
each event. The signal \deltat PDF for $\Bztopippim$ is given by
\begin{equation}
\begin{split}
f^{\pm}_{k}(\deltat_{\rm meas}) = \frac{e^{-|\deltat|/\tau}}{4\tau}
           \Bigl\{ &( 1 \mp \Delta w )  \\
                   & \pm ( 1-2w_{k} ) \bigl[  \spipi \sin{(\deltamd\deltat)} -  \cpipi\cos{(\deltamd\deltat)}    
                   \bigr]   
           \Bigr\} \otimes R(\deltat_{\rm meas} - \deltat),
\end{split}
\end{equation} 
where $f^{+}_{k}$ ($f^{-}_{k}$) indicates a \Bz (\Bzb) flavor tag, and
where index $k$ is the tagging category.  The resolution function
$R(\deltat_{\rm meas} - \deltat)$ for signal candidates is a sum of three Gaussians,
identical to the one described in Ref.~\cite{BaBarsin2beta}, with
parameters determined from a fit to the \Bflav\ sample (including
events in all seven tagging categories).  The background $\deltat$
distribution is modeled as the sum of three Gaussian functions, where
the common parameters used to describe the background shape for all
tagging categories are determined simultaneously with the \CP
parameters in the maximum likelihood fit.

The ML fit includes eight components: \B\ decays and background with
the final states
$\pip\pim$, $\Kp\pim$, $\Km\pip$, and $\Kp\Km$.  The component
coefficients for $K\pi$ are parametrized as
$n_{\Kpm\pimp}=n_{K\pi}\left(1\mp \akpi\right)/2$, where $\akpi$ is
the direct \CP-violating asymmetry.  All other coefficients are the
product of the fraction of events in each tagging category, taken from
\Bflav\ events, and the event yield.  The background PDFs
are a threshold function for \mes and a polynomial for \de.  The \fish
PDF is a double Gaussian for the background and an asymmetric Gaussian
for the signal.  All
background PDF parameters are allowed to float in the ML fit.

\subsection{\boldmath \Bztopizpiz}
\label{sec:pizpizMethod}

\Bztopizpiz events are identified with an ML fit to the variables \mes,
\de, and \fish.  For \Bztopizpiz, we require $\mes>5.20\gevcc$ and
$|\de| < 0.2~\gev$.  Tails in the EMC response produce a correlation
between \mes and \de, so a two-dimensional PDF, derived from Monte
Carlo (MC) simulation, is used.  The PDF for \fish is a step
function, with parameters taken from MC.  \Bflav\ data is used to
verify that the MC accurately reproduces the \fish distribution.  The
\qq background PDFs are a threshold function for \mes, a polynomial
for \de, and a step function for \fish.  All \qq background PDF
parameters are allowed to float in the ML fit.

The decays \Bptorhoppiz and $\Bztokzpiz (\KS\to\piz\piz)$ add $52 \pm
6.8$ background events to \Bztopizpiz and are included as an
additional fixed component in the ML fit.  We model these \B
backgrounds with a two-dimensional PDF to describe \mes and \de and
with a step function for \fish, all taken from MC simulation.

The time-integrated \CP asymmetry is measured by the \B-flavor
tagging described previously.  The fraction of events in each tagging
category is also constrained to the fractions determined from
MC simulation.  The PDF coefficient for the \Bztopizpiz signal is
given by the expression
\begin{equation}
n_{\piz\piz, k} = \frac{1}{2} f_{k} N_{\piz\piz} \Bigl\{ 1 - s_j
  (1-2\chi)(1-2w_k) \cpizpiz \Bigr\},
\end{equation}
where $f_k$ is the fraction of events in tagging category $k$,
$N_{\piz\piz}$ is the number of \B decays, $\chi = 0.184 \pm
0.004$~\cite{pdg} is the time-integrated mixing probability, and $s_j=+1(-1)$
when the \Btag\ is a \Bz (\Bzb).

\subsection{\boldmath\Btopipiz and \Btokpiz}
\label{sec:hpizMethod}

\Btopipiz and \Btokpiz events are identified in an ML fit to the
variables \mes, \de, \fish, and \thetac.  We require \B candidates to
satisfy $\mes>5.22\gevcc$ and $-0.11 < \de < 0.15~\gev$. The tighter
requirement on \de serves to remove \B-decay backgrounds.  The
treatment of \de and the use of the Cherenkov angle for kaons and
pions is identical to that in \Bztopipi.  The \fish distribution is
also described with a step function, with parameters taken from
MC simulation.  The \qq background PDFs are a threshold
function for \mes, a polynomial for \de, and a step function for
\fish.  All \qq background PDF parameters are floating in the ML fit.

Several charmless \B decays with both a high momentum charged track
and a \piz or photon appear as background; significant contributions, as determined with MC,
come from the following decay modes with the estimated number of events provided 
in parentheses: \Bztorhoppim
($32\pm3$ events), \Bptorhoppiz ($20\pm 2$), $\B \to X_{s} \gamma$ ($5.8\pm 0.6$),
$\Btokzpi (\KS \to \piz\piz)$ ($5.0\pm 0.3$), and \Bztorhopkm ($3.3\pm0.5$).  These
events, as well as their measured \CP asymmetry, are included as an
additional fixed component in the ML fit.

The PDF coefficient for \Btopipiz and \Btokpiz signal is
given by the expression
\begin{equation}
n_i = \frac{1}{2} N_{i}  ( 1 - q_j \mathcal{A}_i ),
\end{equation}
where $\mathcal{A}_i$ is the direct \CP asymmetry and $q_j$ is the
charge of the \B candidate.

\section{RESULTS AND SYSTEMATIC UNCERTAINTIES}
\label{sec:Physics}

\begin{figure}[!tbhp]
\begin{center}
  \includegraphics[width=0.49\linewidth]{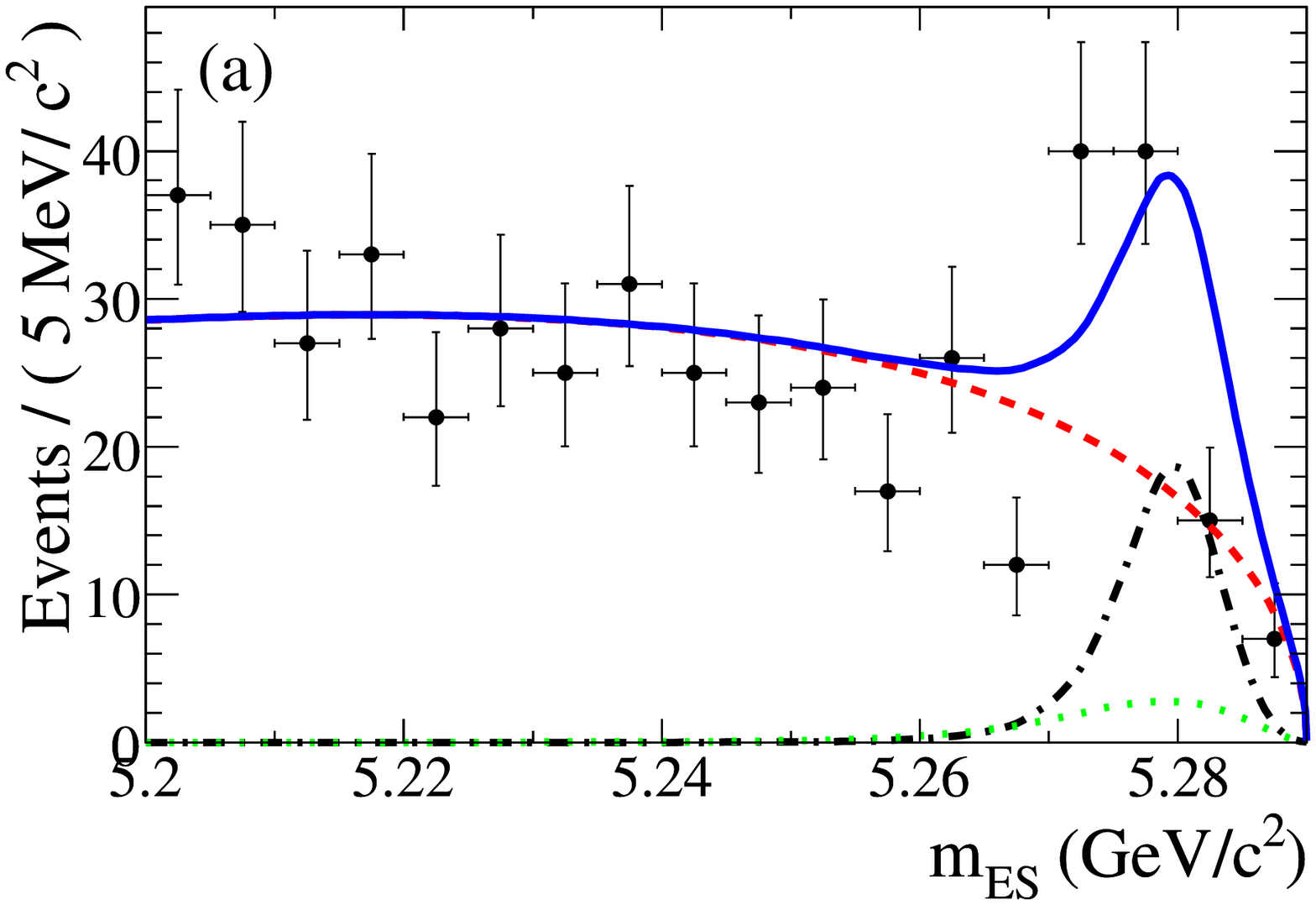}
  \includegraphics[width=0.49\linewidth]{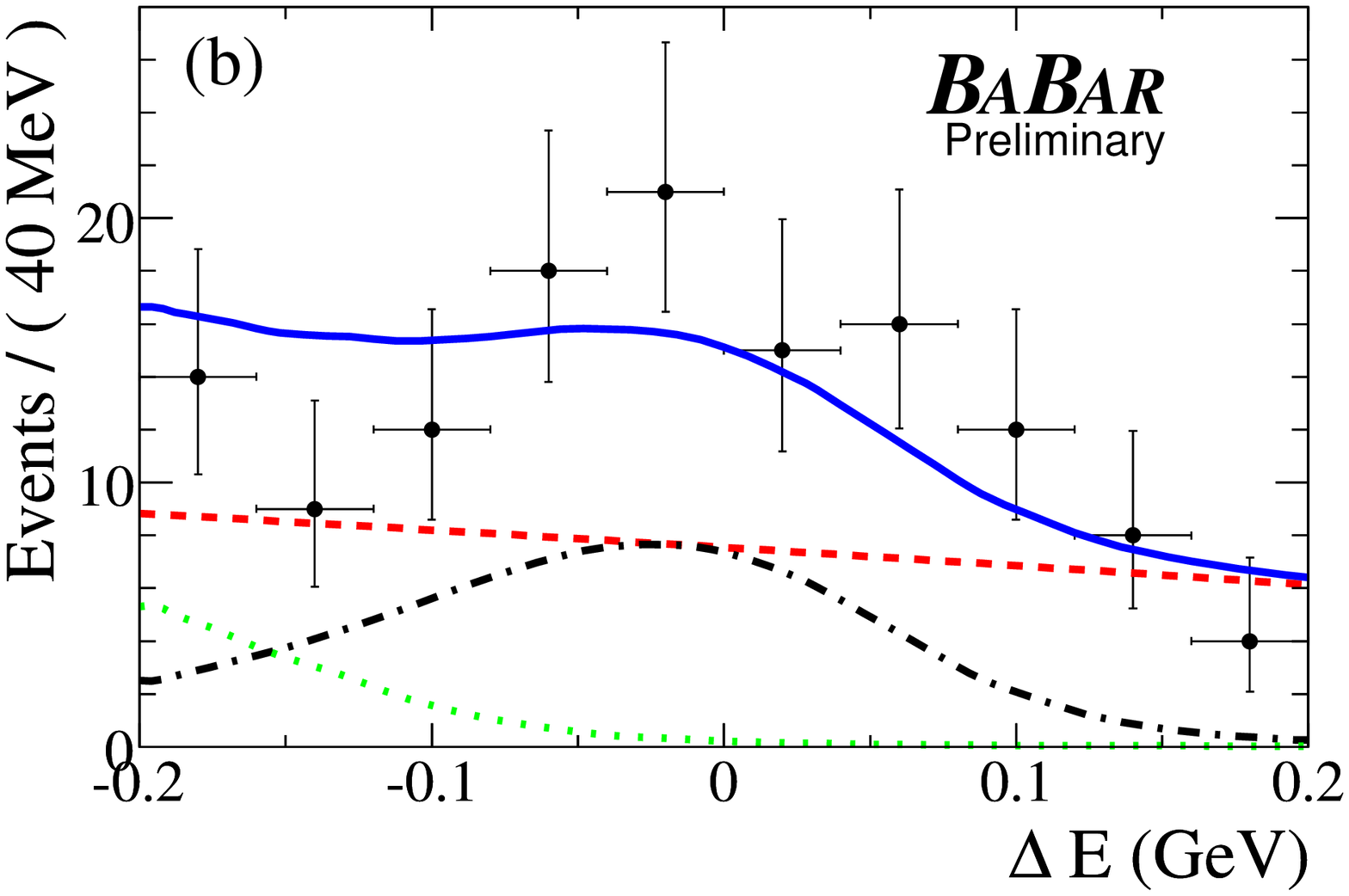}
  \includegraphics[width=0.49\linewidth]{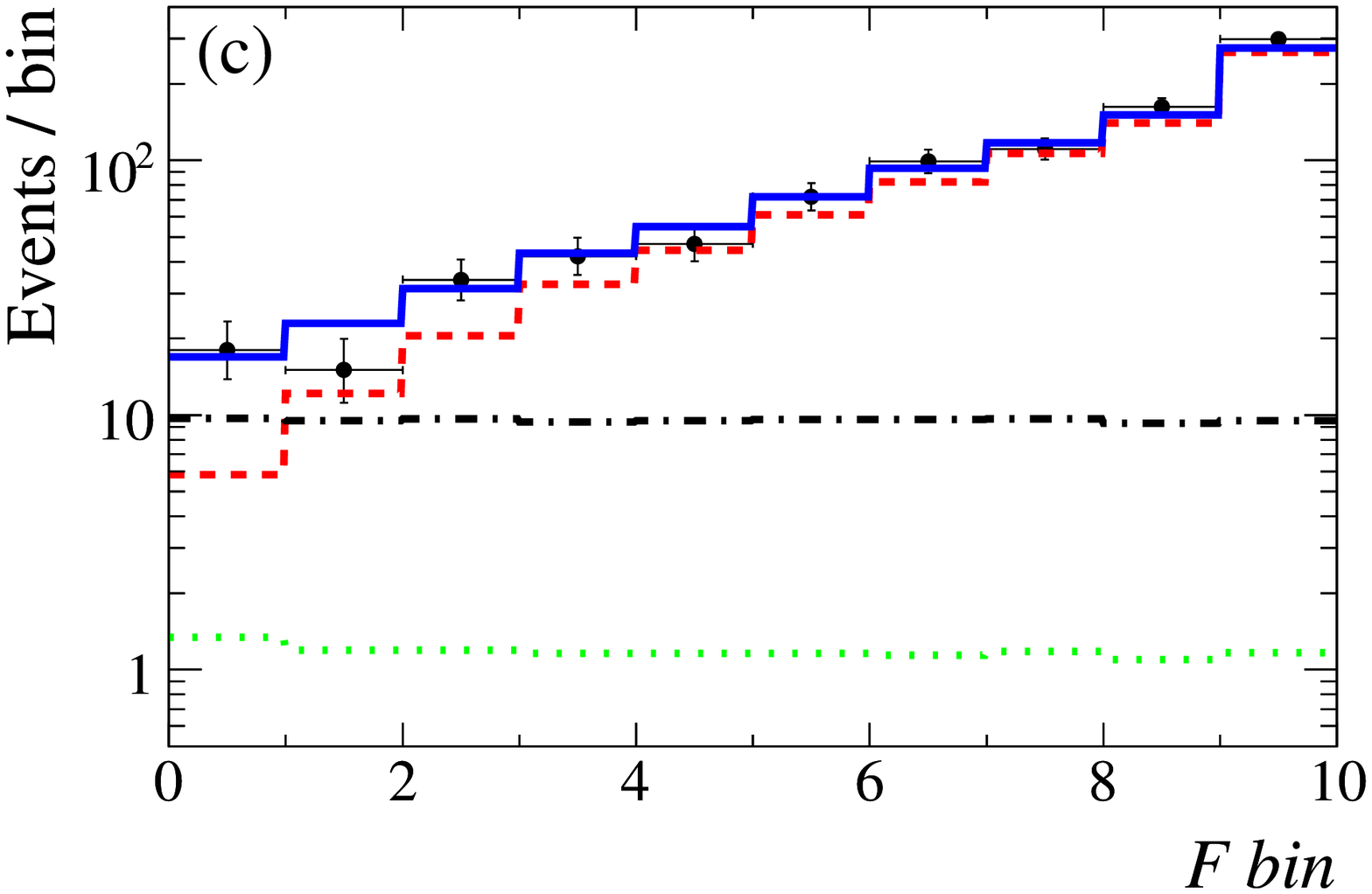}
  \includegraphics[width=0.49\linewidth]{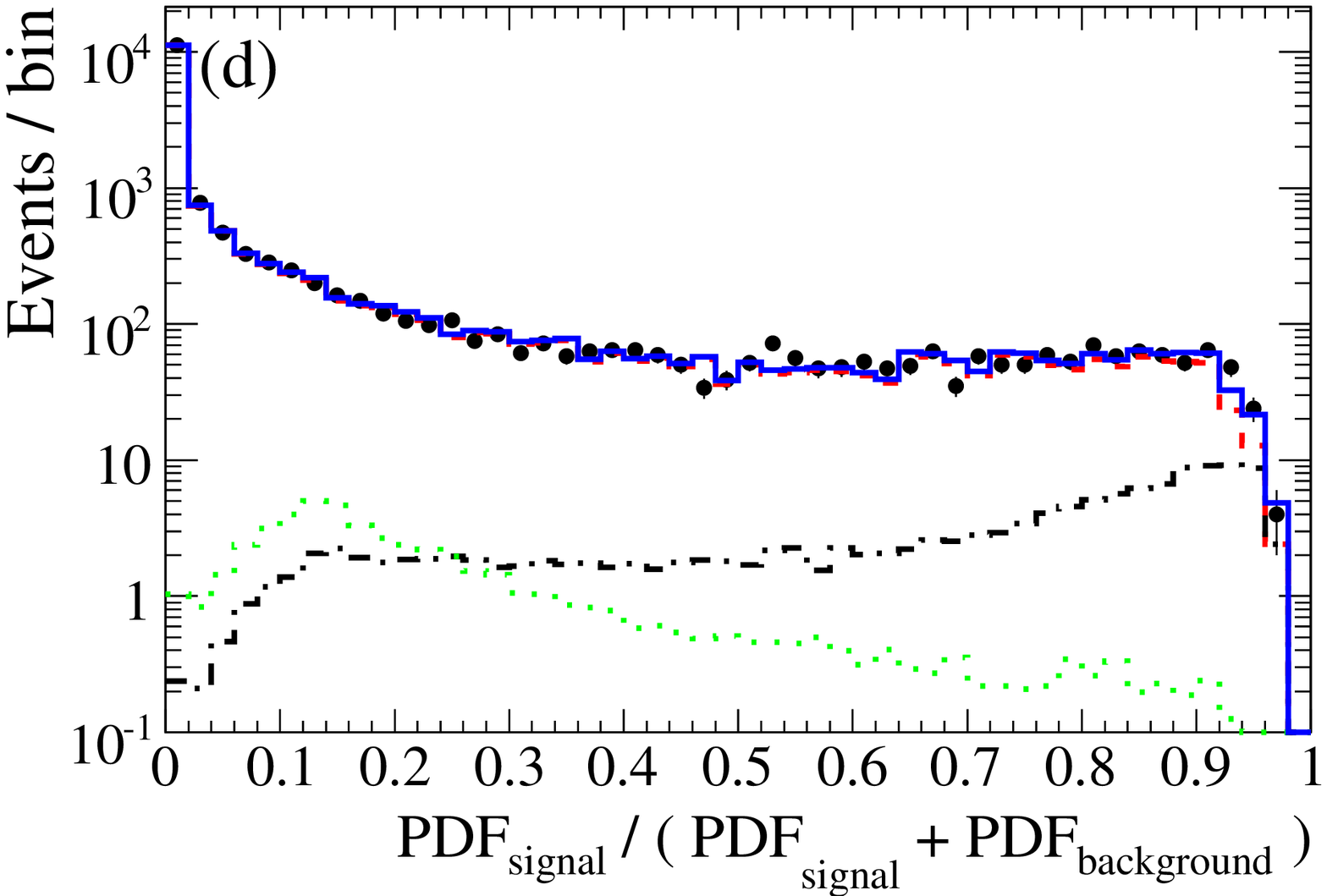}
\caption{ The  distributions of (a) \mes, (b) \de, and (c) Fisher
discriminant \fish for \Bztopizpiz
candidates that satisfy an optimized
requirement on the signal probability, based on
all variables except the one being plotted.  
The projections contain 
27\%, 30\% and 68\% 
of the signal, 
20\%, 21\% and 23\%
of the $\rho\piz$
background, and 
2.8\%, 0.047\% and 4.8\% 
of the continuum background,
for \mes, \de, and \fish, respectively.
The PDF projections are shown as a dashed
line for \qqbar background, a dotted line for \Btorhopiz and \Bztokzpiz, and a
dashed-dotted line for \Bztopizpiz signal.  The solid line shows the
sum of all PDF projections.  The PDF projections are scaled by
the expected fraction of events passing the probability-ratio
requirement. Also shown (d) is the ratio of the PDF for signal to the
PDF for signal plus background comparing data (points) to the
components of the PDF model.}
\label{fig:pizpiz}
\end{center}
\end{figure}

\begin{figure}[!tbhp]
\begin{center}
\includegraphics[width=0.49\linewidth]{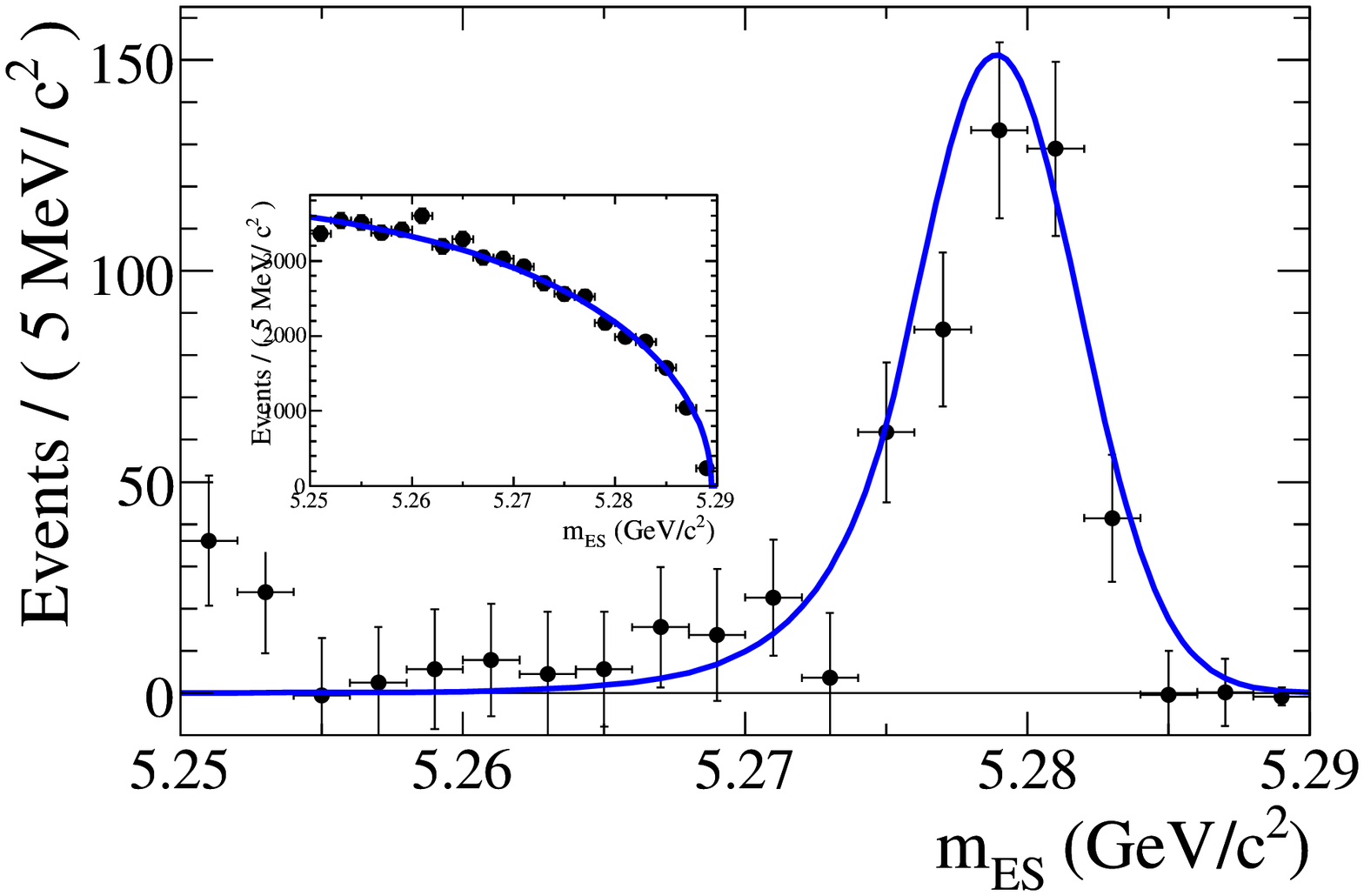}
\includegraphics[width=0.49\linewidth]{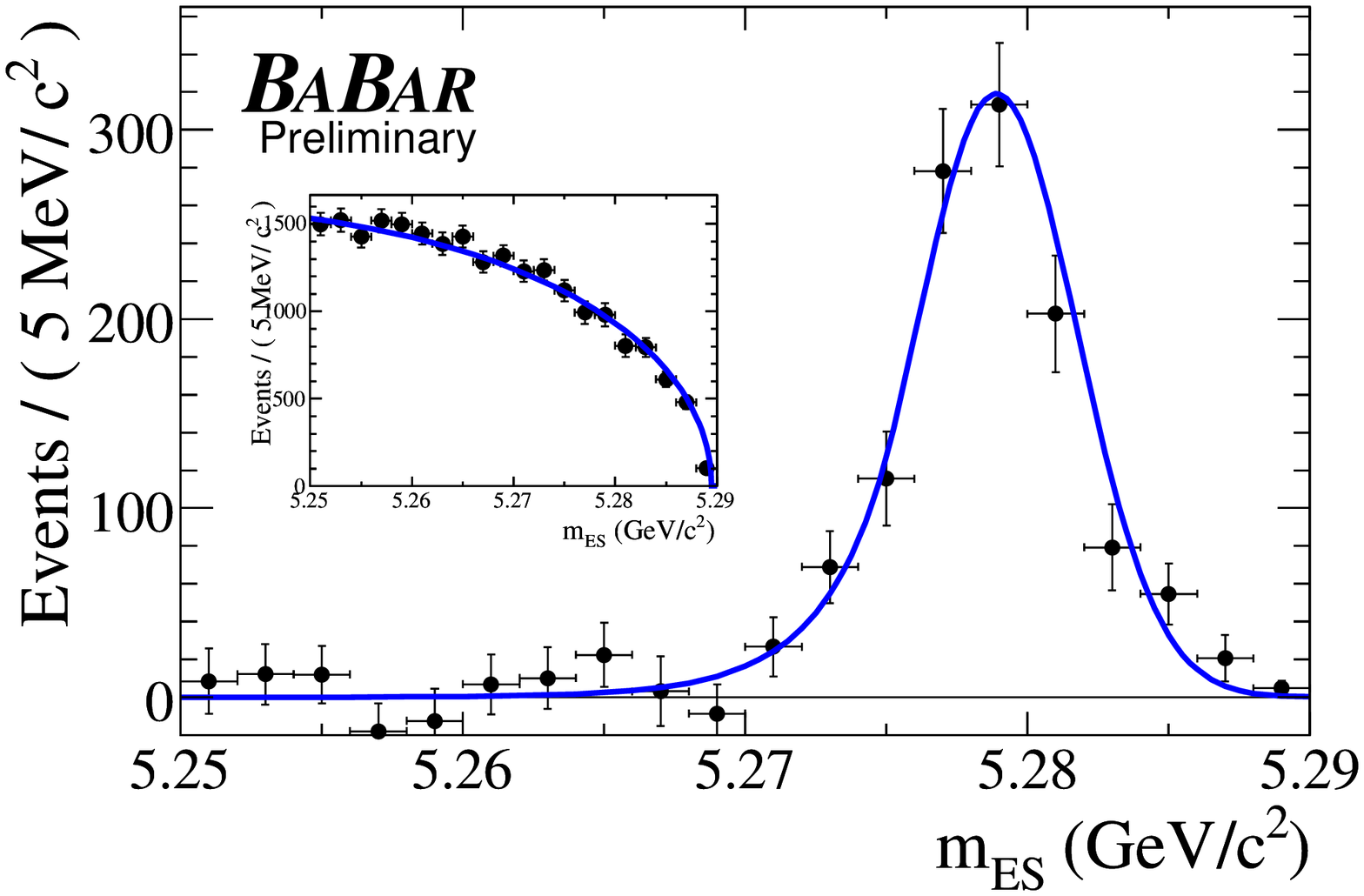}
\includegraphics[width=0.49\linewidth]{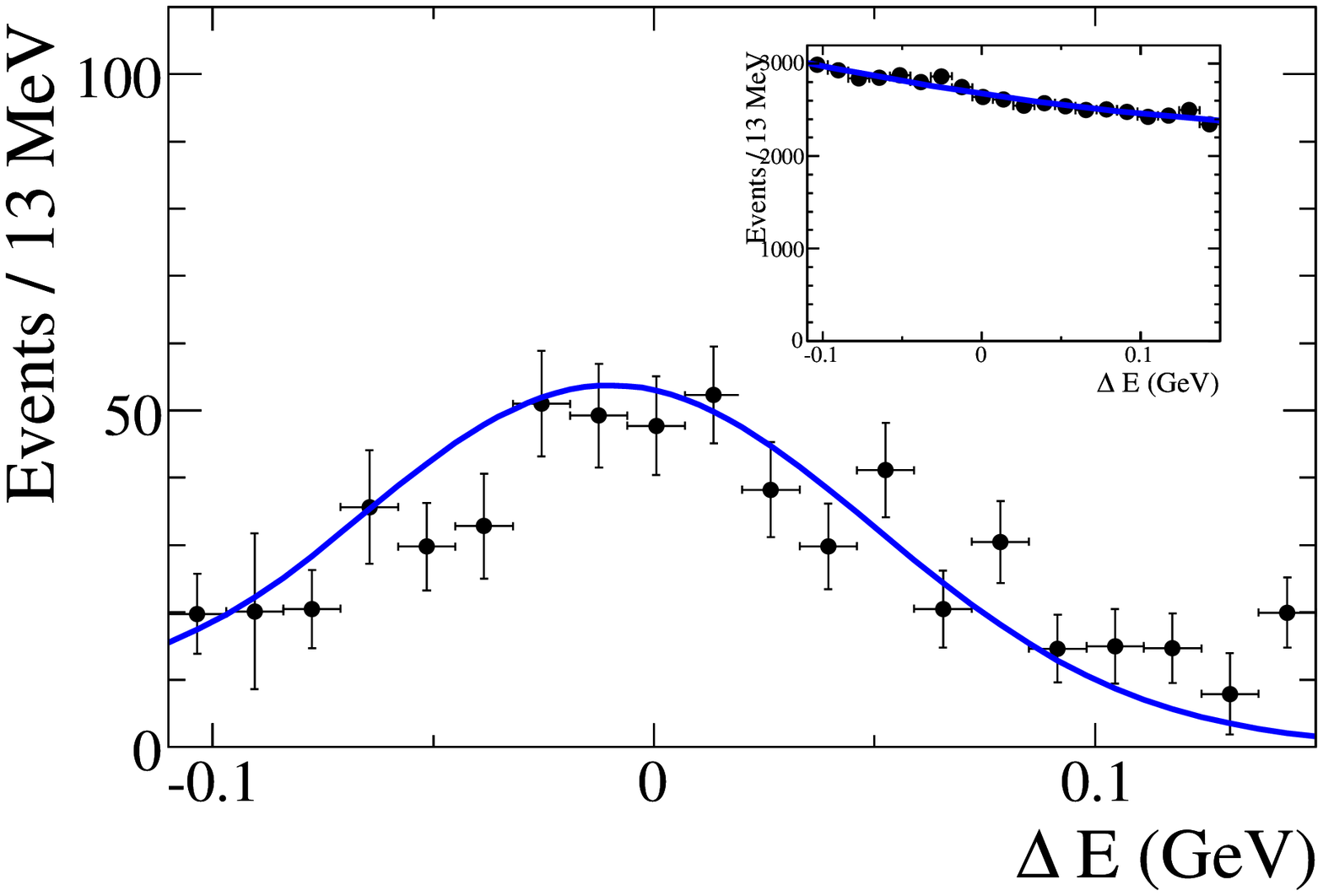}
\includegraphics[width=0.49\linewidth]{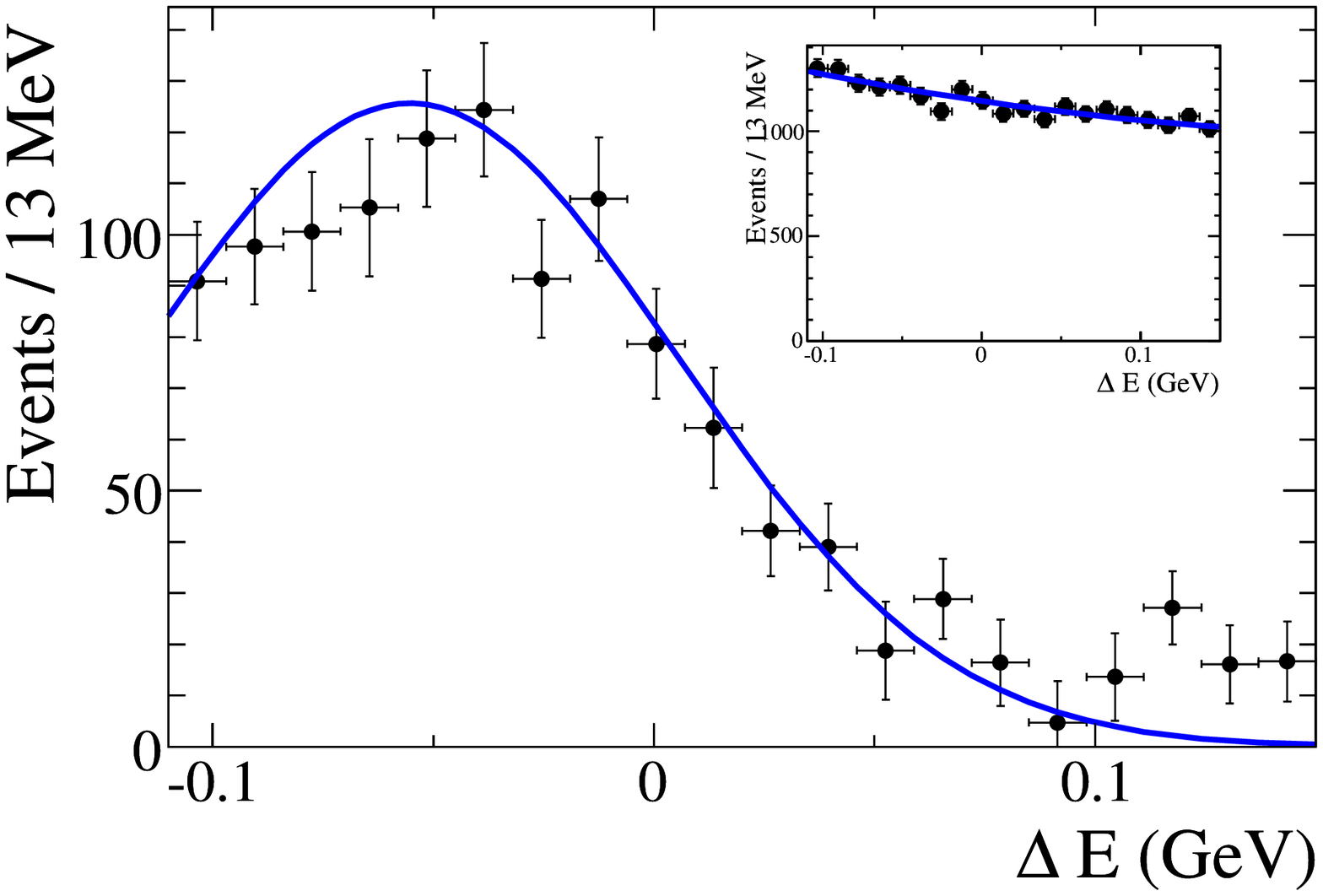}
\includegraphics[width=0.49\linewidth]{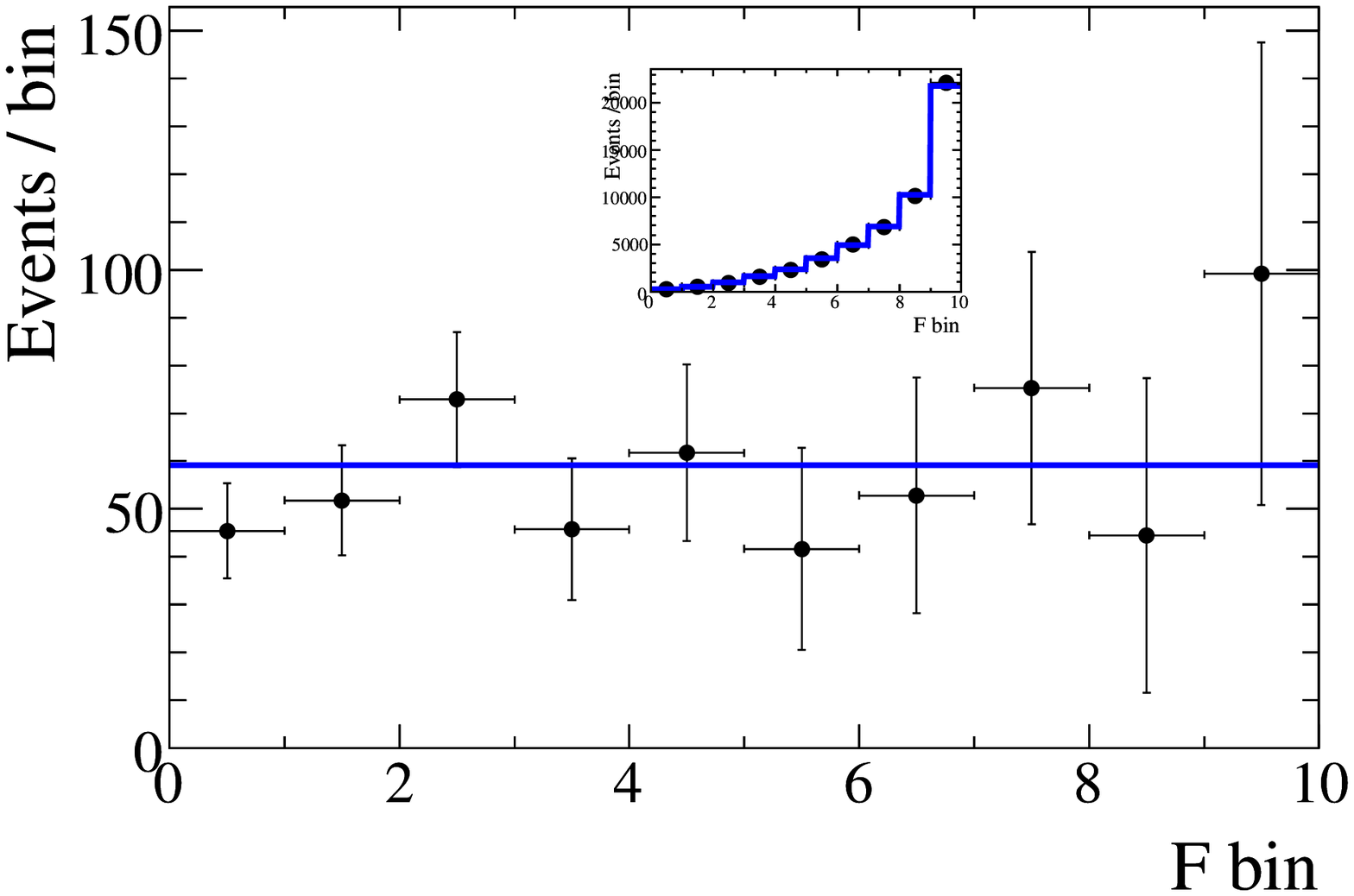}
\includegraphics[width=0.49\linewidth]{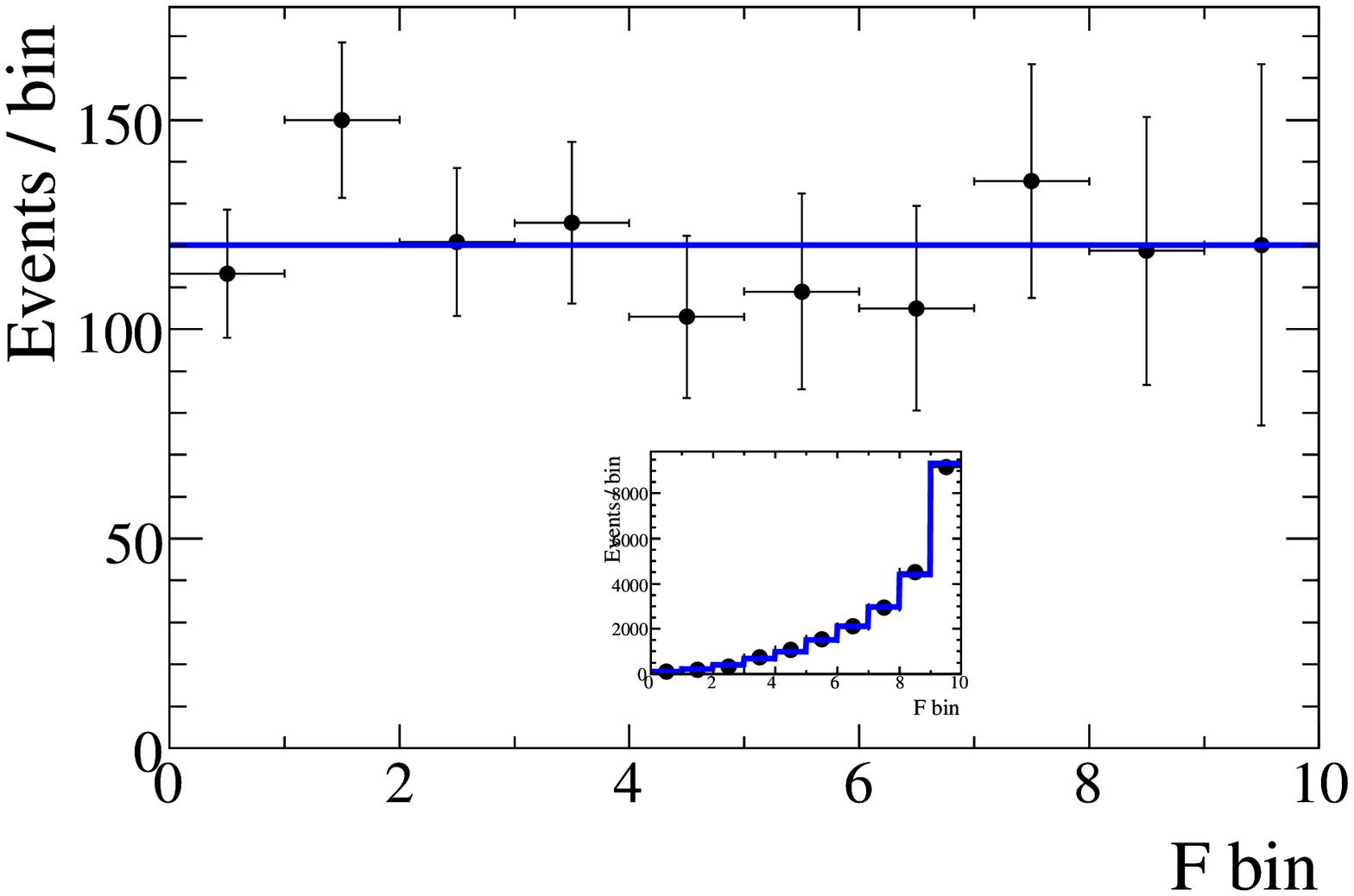}
\end{center}
\vspace{0.2cm}
\caption{
  The distributions and PDF projections of \mes (top), \de (middle),
  and Fisher discriminant \fish (bottom), for \Btopipiz (left) and \Btokpiz
  (right) candidates. The main plots show the signal data (points)
  and PDF (line) after event weighting and background subtraction
  with the method described in~\cite{splots}; the method
  uses all variables except the one being plotted.  The insets show
  the corresponding distributions for background. }
\label{fig:hpiz}
\end{figure}

Results from the ML fits for the \Bztopizpiz and \Btohpiz decay modes are
summarized in Table~\ref{tab:resultsA}. Distributions of \mes, \de,
and \fish for \Bztopizpiz are in Fig.~\ref{fig:pizpiz}, where a
signal-enhanced subset of the data is shown.  The \Btohpiz
data are shown in Fig.~\ref{fig:hpiz}.  With a large signal in both
decay modes, we show weighted and background-subtracted
plots~\cite{splots} for the \Btopipiz and \Btokpiz signal.  The
same technique is used to display the \qq background as well.

The uncertainty in the efficiency for the \Bztopizpiz decay mode is
dominated by a $3\%$ systematic uncertainty per \piz, estimated from a
study of $\tau \to \pi\piz\nu_{\tau}$ decays. There is an additional
$3.6\%$ uncertainty from our knowledge of the EMC resolution function,
based on a study of the resolution of the \piz mass and the energy of
the photon in $\epem \to \mu^+\mu^-\gamma$ events.  Systematic
uncertainties involving the ML fit are evaluated by varying the PDF
parameters and refitting the data. The changes in the \mes and \de
signal PDFs are taken from the difference in these quantities in the
\Btohpiz sample between data and MC. The change in the result is taken
as the systematic error. All sources of systematic error for
\Bztopizpiz are listed in Table~\ref{tab:pizpizsyst}.

The largest uncertainties in the \Btohpiz decays arise from
uncertainty in the \mes and \de PDFs, our knowledge of the signal
\fish distribution evaluated from a sample of \Bpm decays, and the 3\%
\piz efficiency uncertainty.  The size of a small bias in the ML fit
is included as a systematic error.  The dominant uncertainty on the
direct \CP asymmetries are taken from the size and error in the
asymmetry fit in the \qq background and the effect of \CP violation in
the \B backgrounds.  The uncertainties in \Btohpiz are summarized in
Table~\ref{tab:hpizsyst}.

\begin{figure}[!tbph]
\begin{center}
  \includegraphics[width=0.49\linewidth]{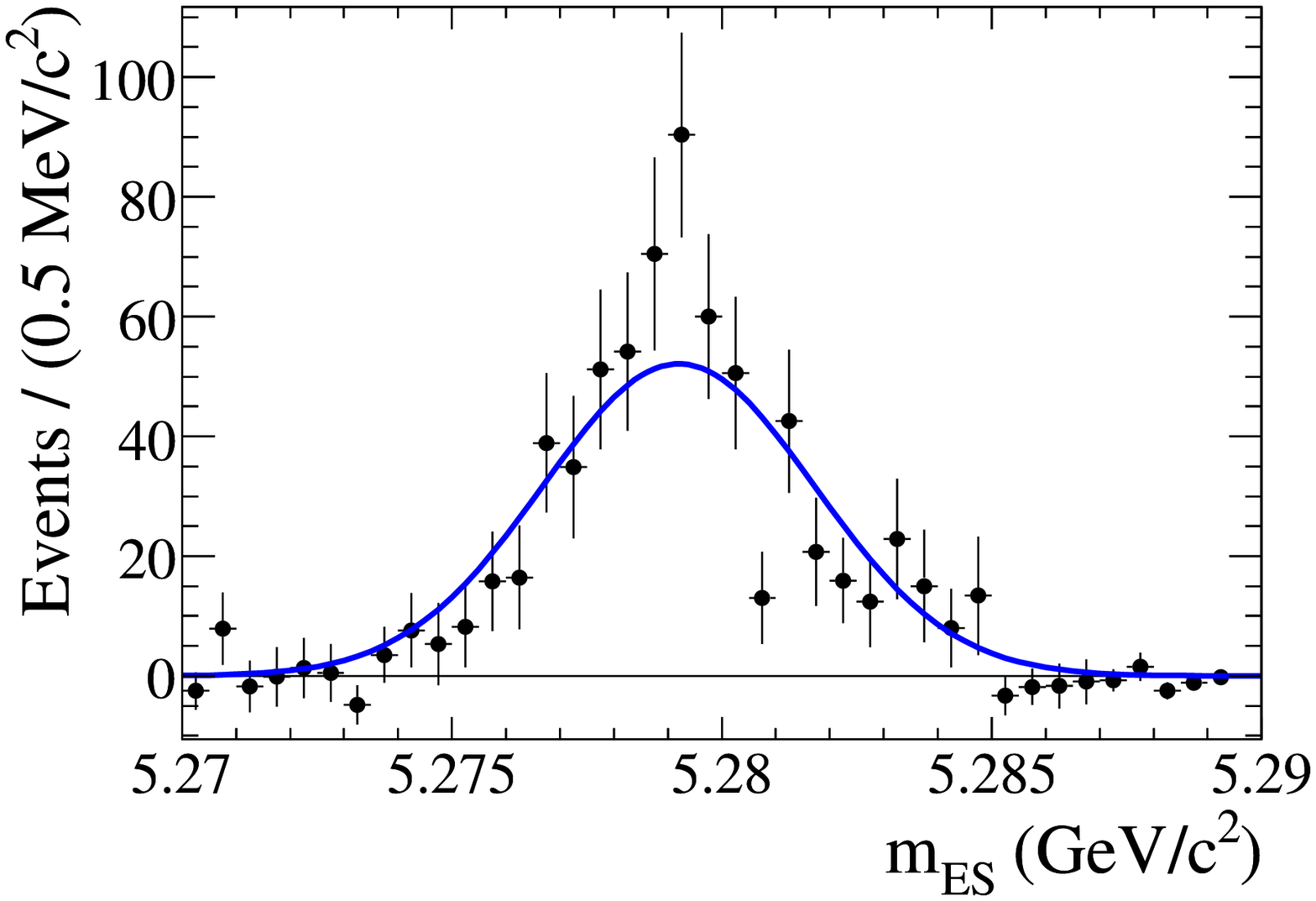}
  \includegraphics[width=0.49\linewidth]{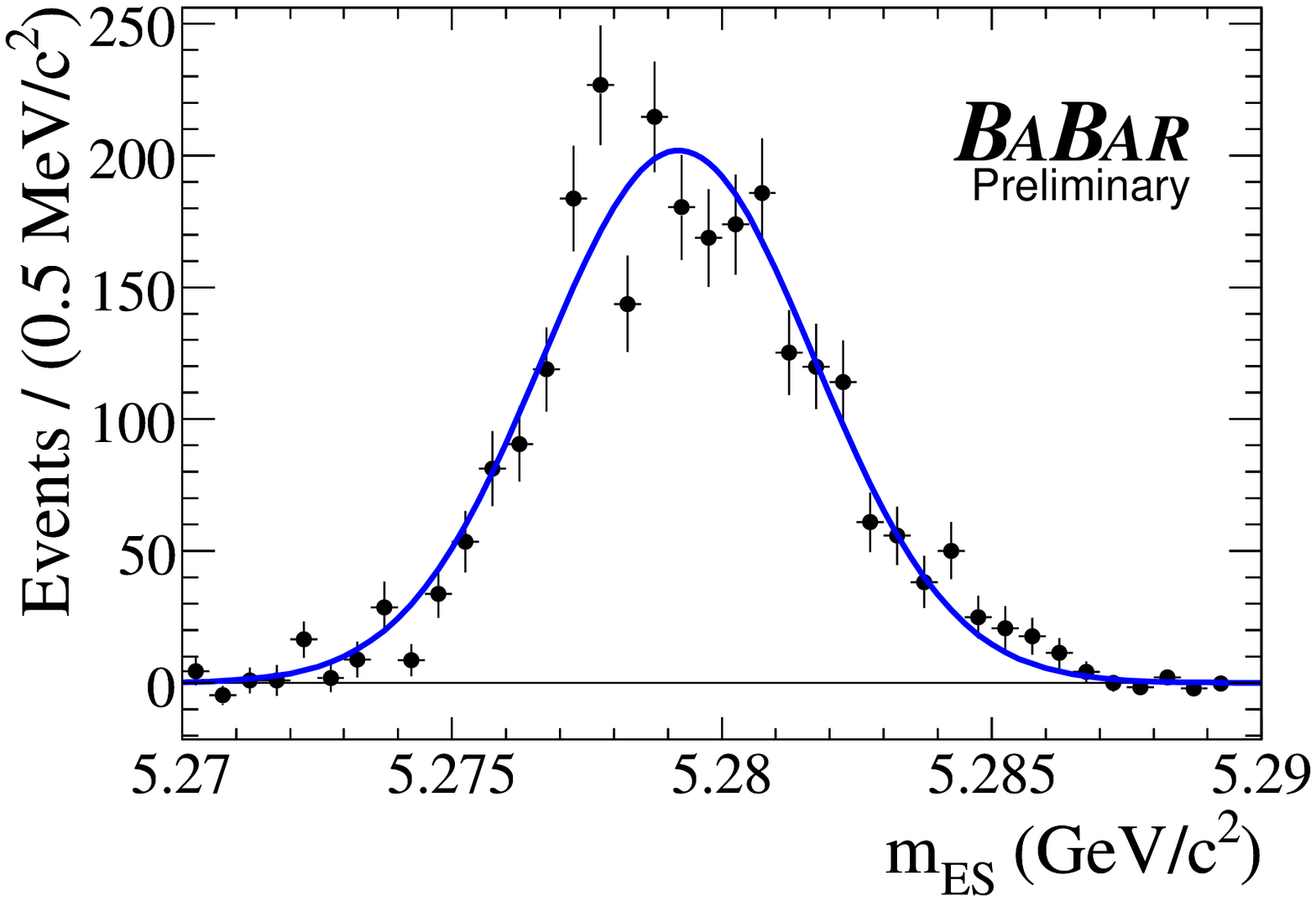}
  \includegraphics[width=0.49\linewidth]{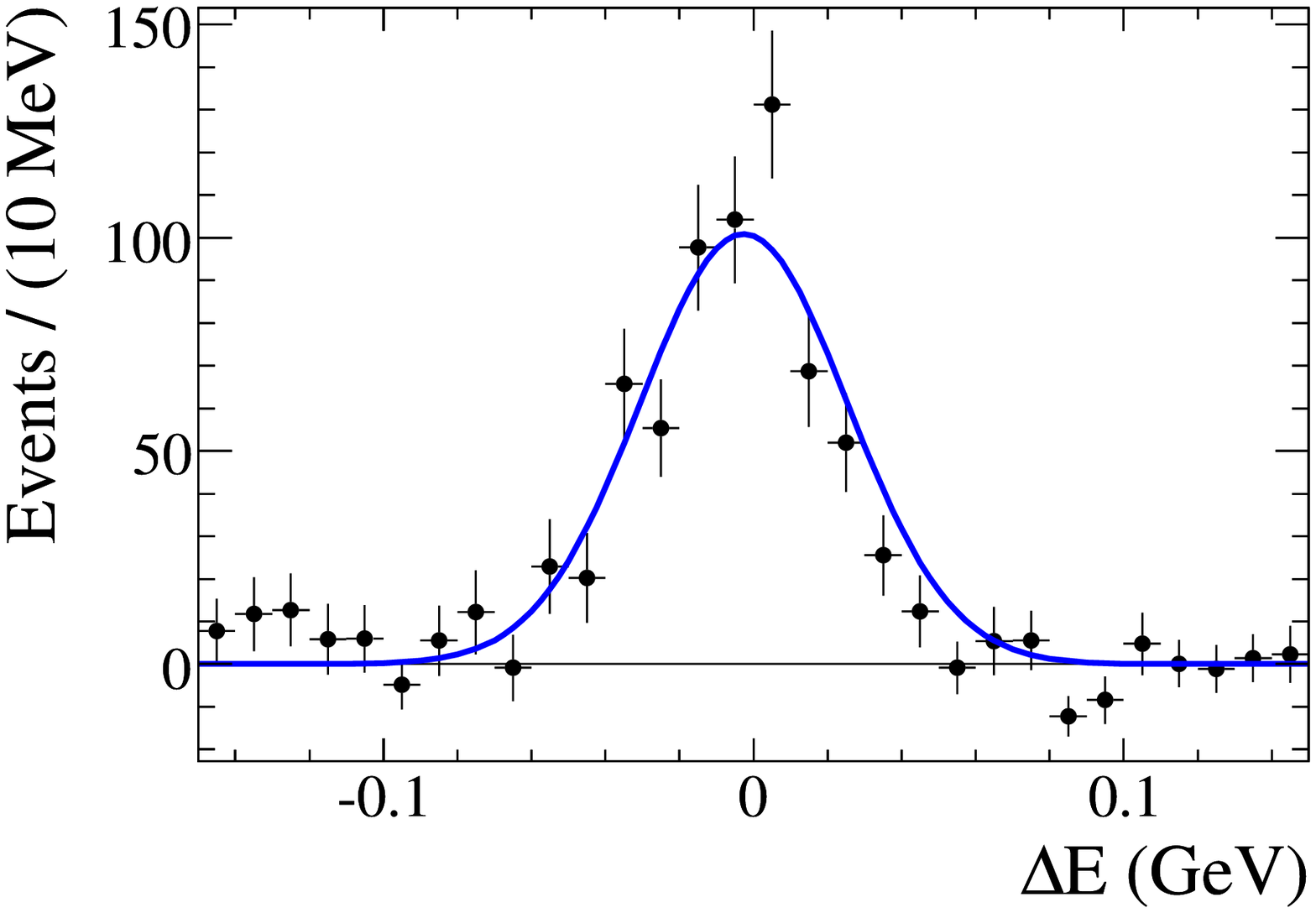}
  \includegraphics[width=0.49\linewidth]{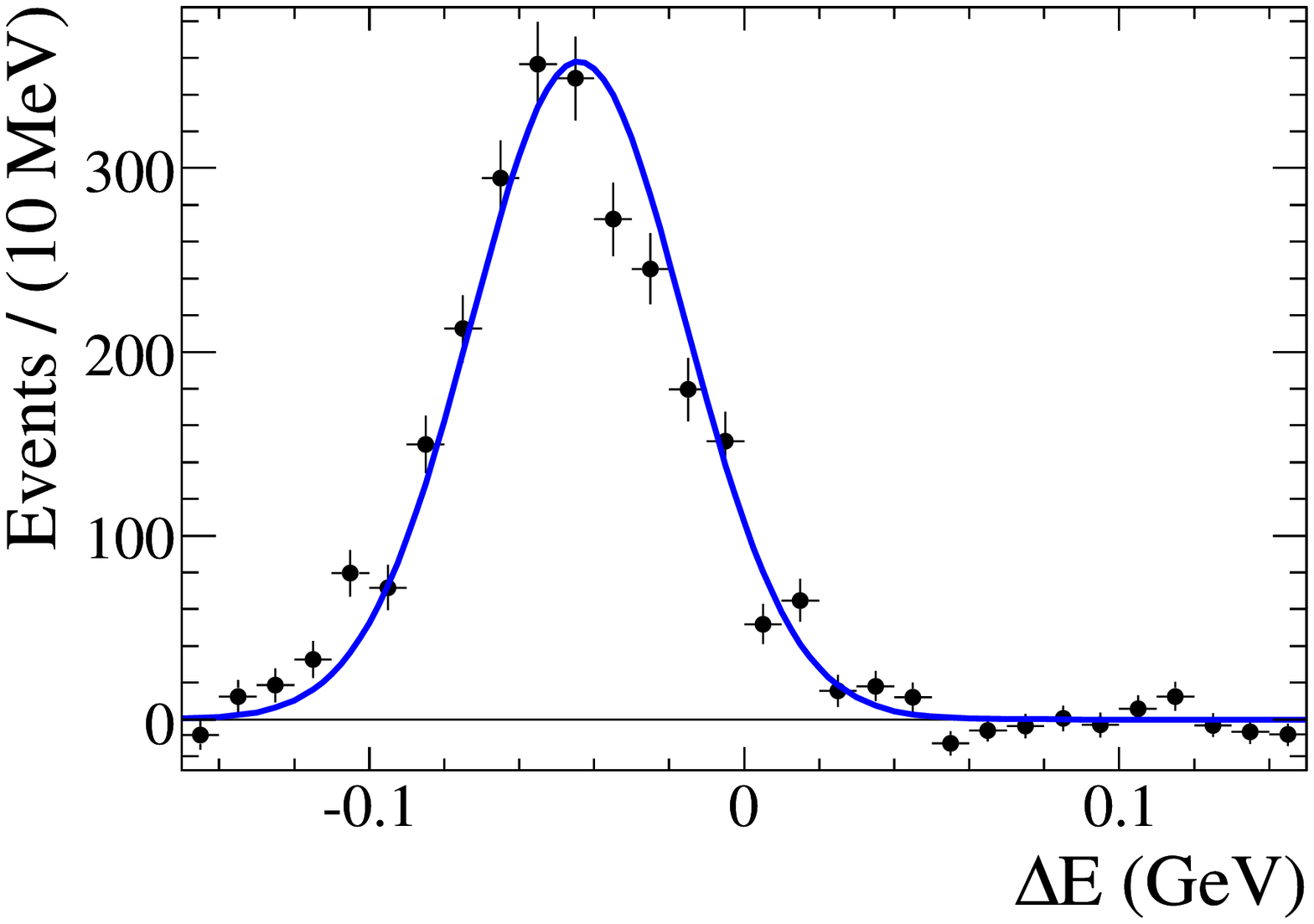}
  \includegraphics[width=0.49\linewidth]{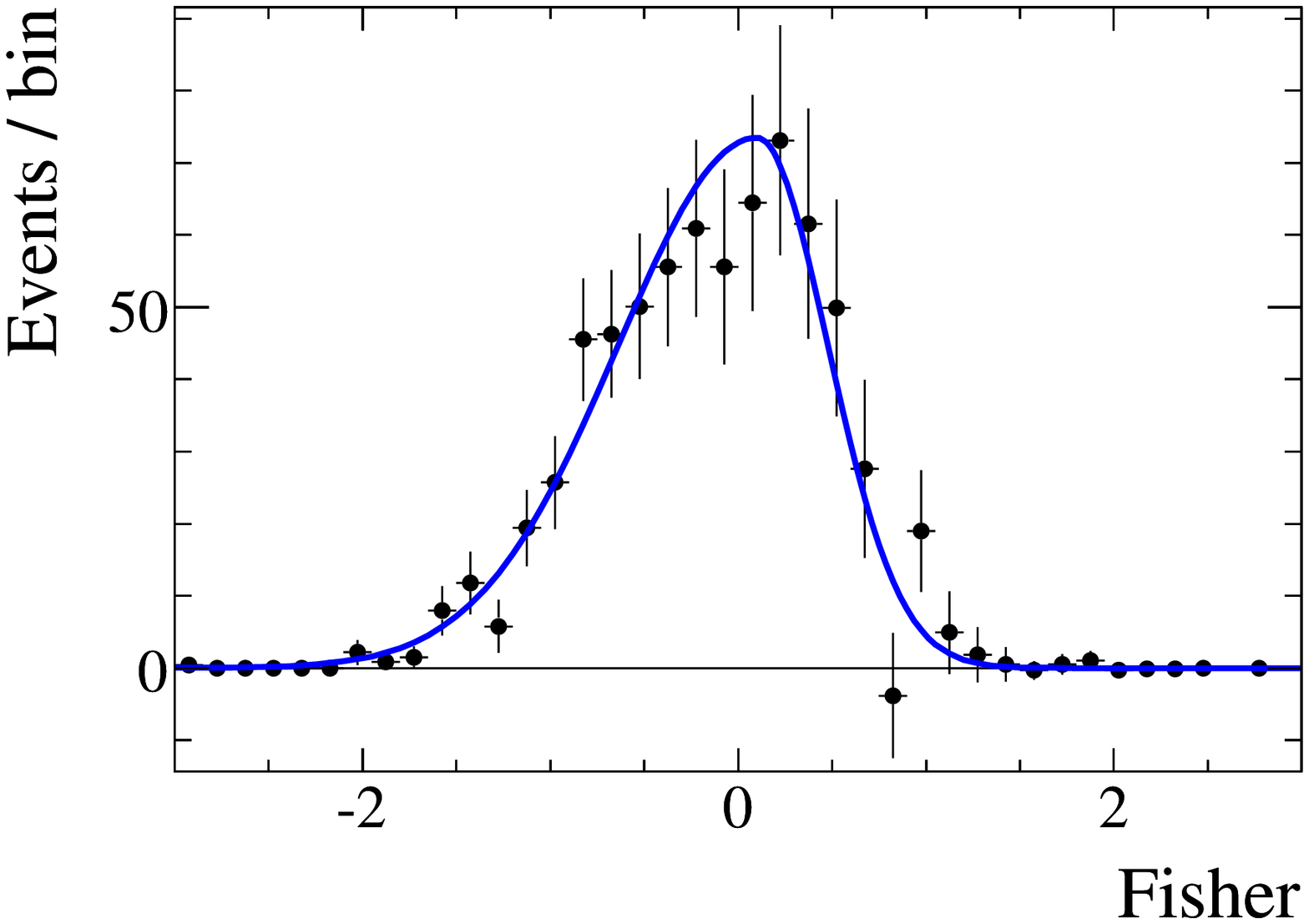}
  \includegraphics[width=0.49\linewidth]{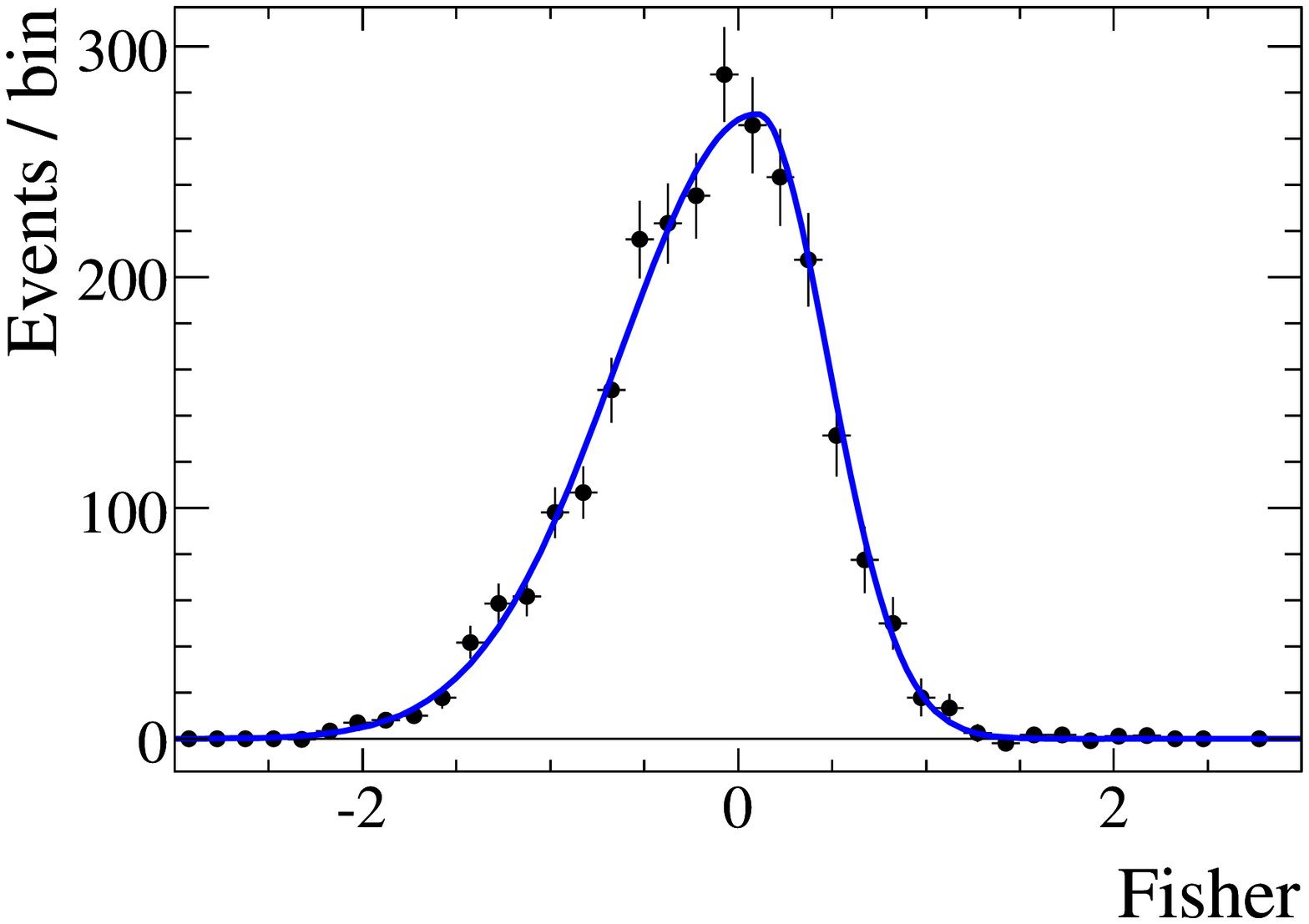}
\caption{ The  background-subtracted distributions of 
\mes (top), \de (middle), and Fisher discriminant 
\fish (bottom) for signal $\Bz\to\pip\pim$ (left) and $\Bz\to\Kp\pim$ (right) 
candidates in the data.  The curves represent the PDFs used in 
the fit.}
\label{fig:hhVarSig}
\end{center}
\end{figure}

\begin{figure}[!tbph]
\begin{center}
  \includegraphics[width=0.32\linewidth]{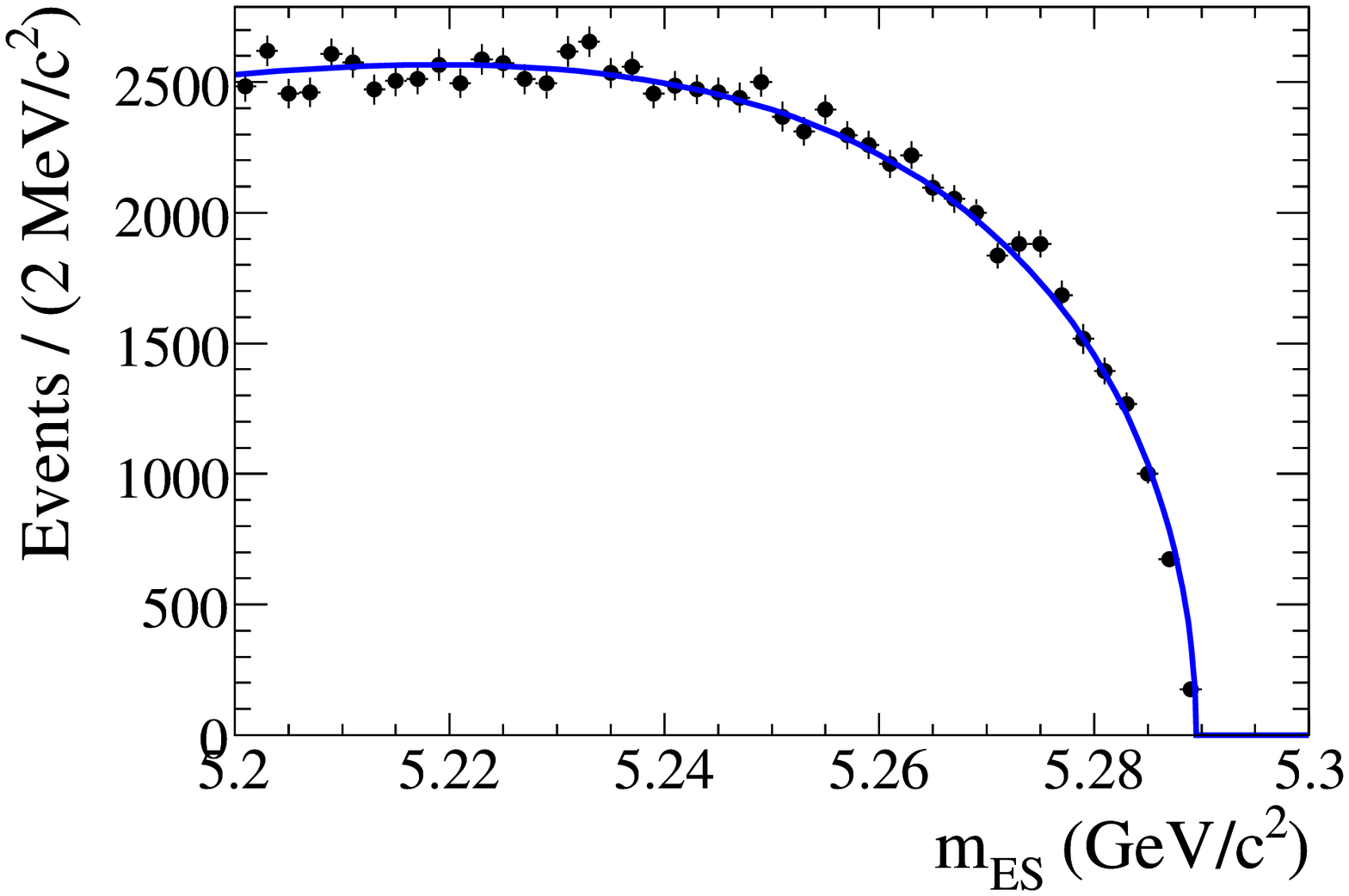}
  \includegraphics[width=0.32\linewidth]{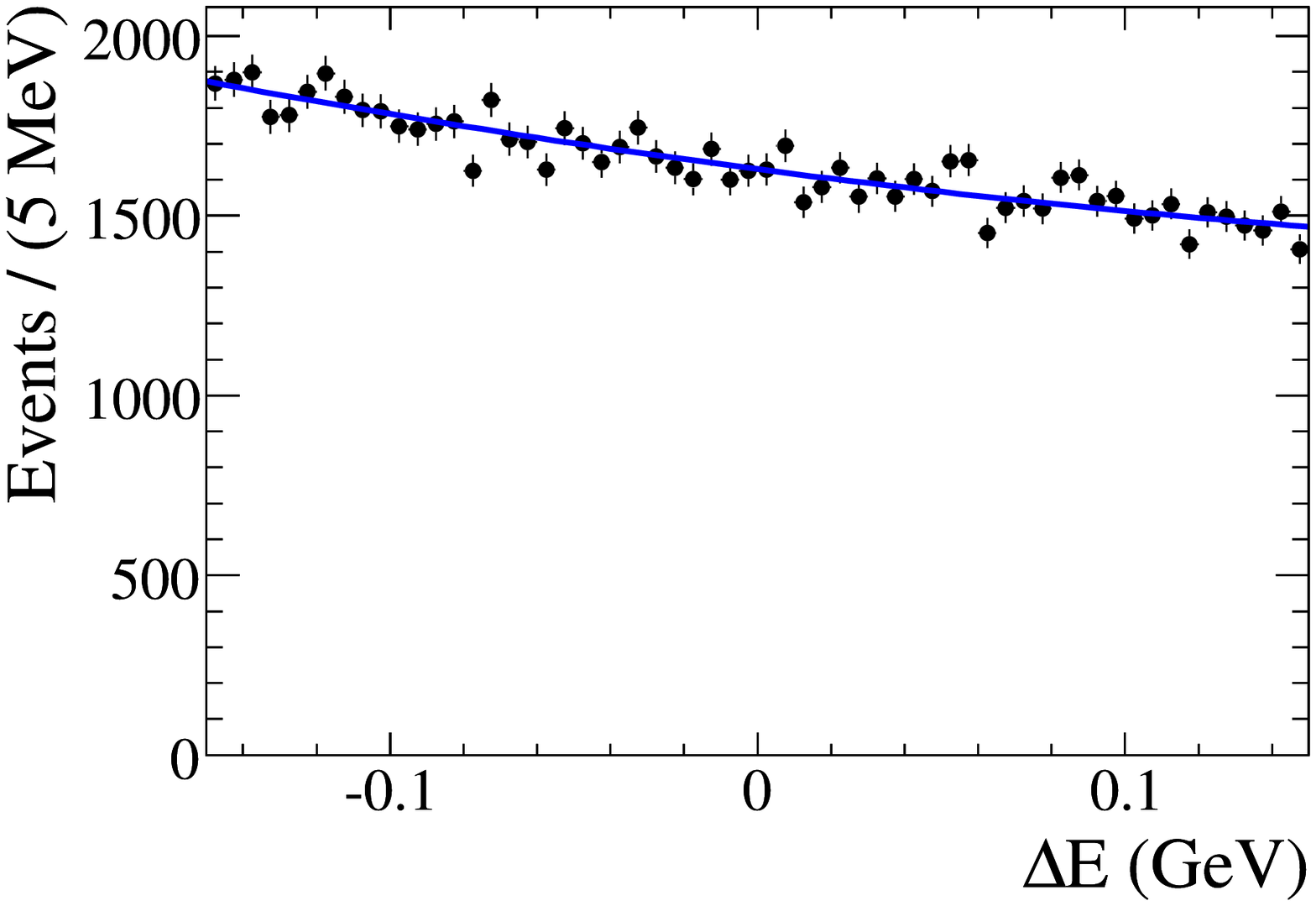}
  \includegraphics[width=0.32\linewidth]{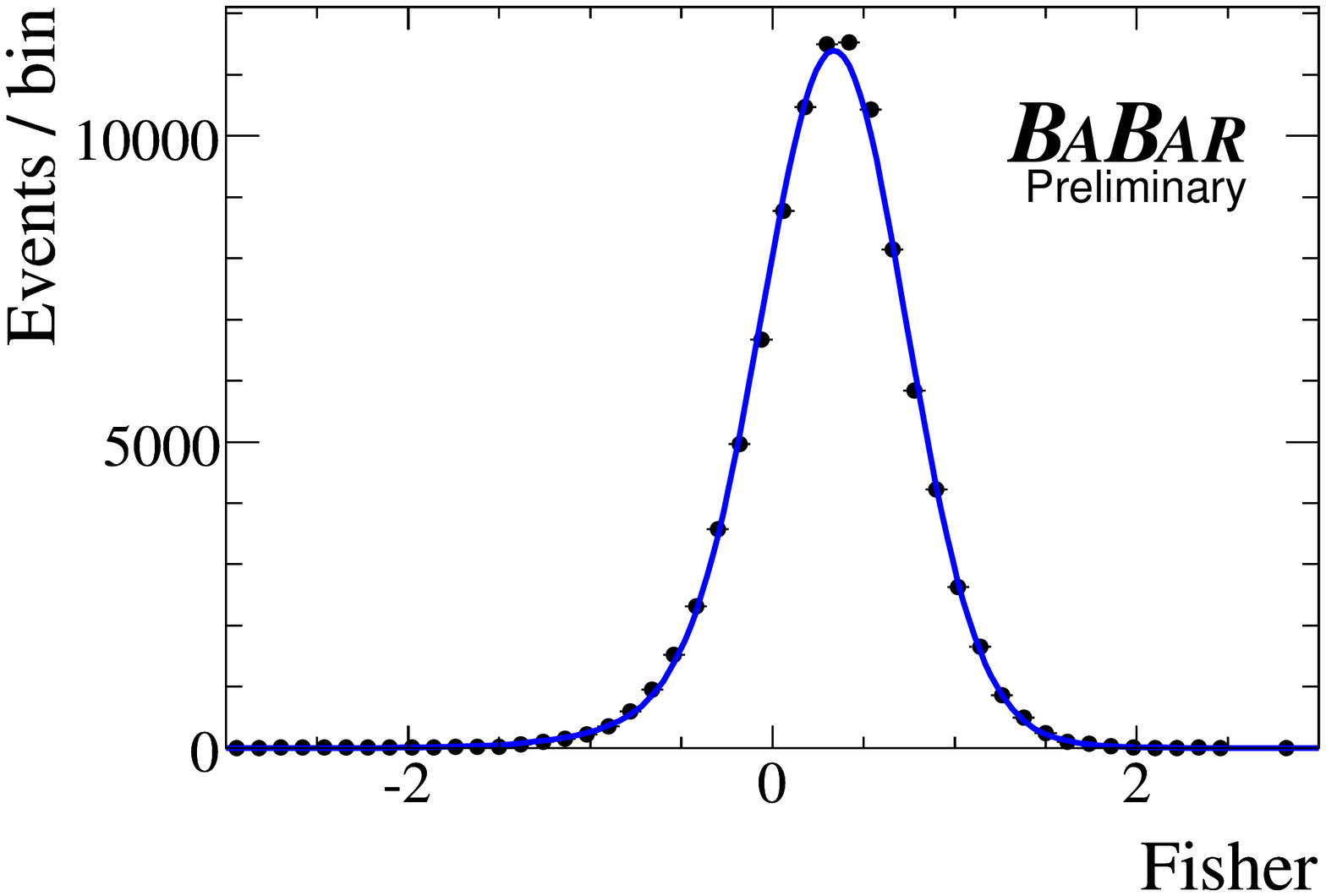}
\caption{ The  signal-subtracted distributions of 
\mes (left), \de (middle), and Fisher discriminant 
\fish (right) for all background $h^+h^{\prime-}$ candidates in the data.  
The curves represent the PDFs used in the fit.}
\label{fig:hhVarBkg}
\end{center}
\end{figure}

\begin{figure}[!tbph]
\begin{center}
  \includegraphics[width=0.69\linewidth]{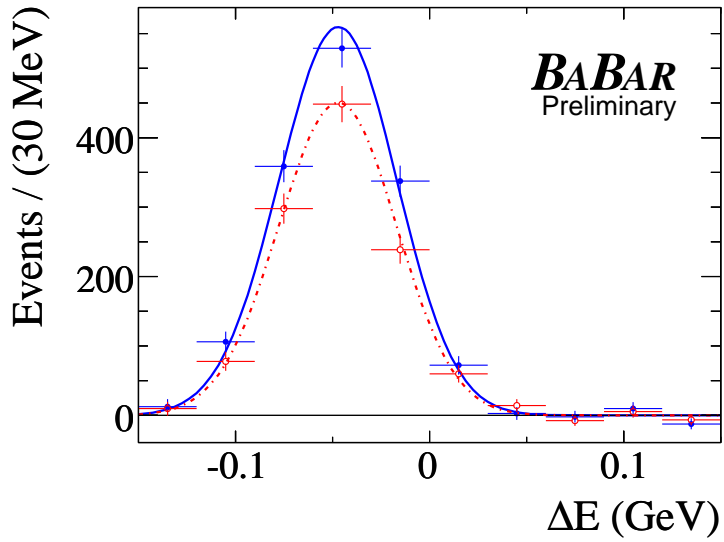}
\caption{ The  background-subtracted distribution of 
$\de$ for signal $\Kpm\pimp$ events, comparing \Bz (solid) and \Bzb decays
(dashed).}
\label{fig:deakpi}
\end{center}
\end{figure}

\begin{figure}[!tbph]
\begin{center}
  \includegraphics[width=0.32\linewidth]{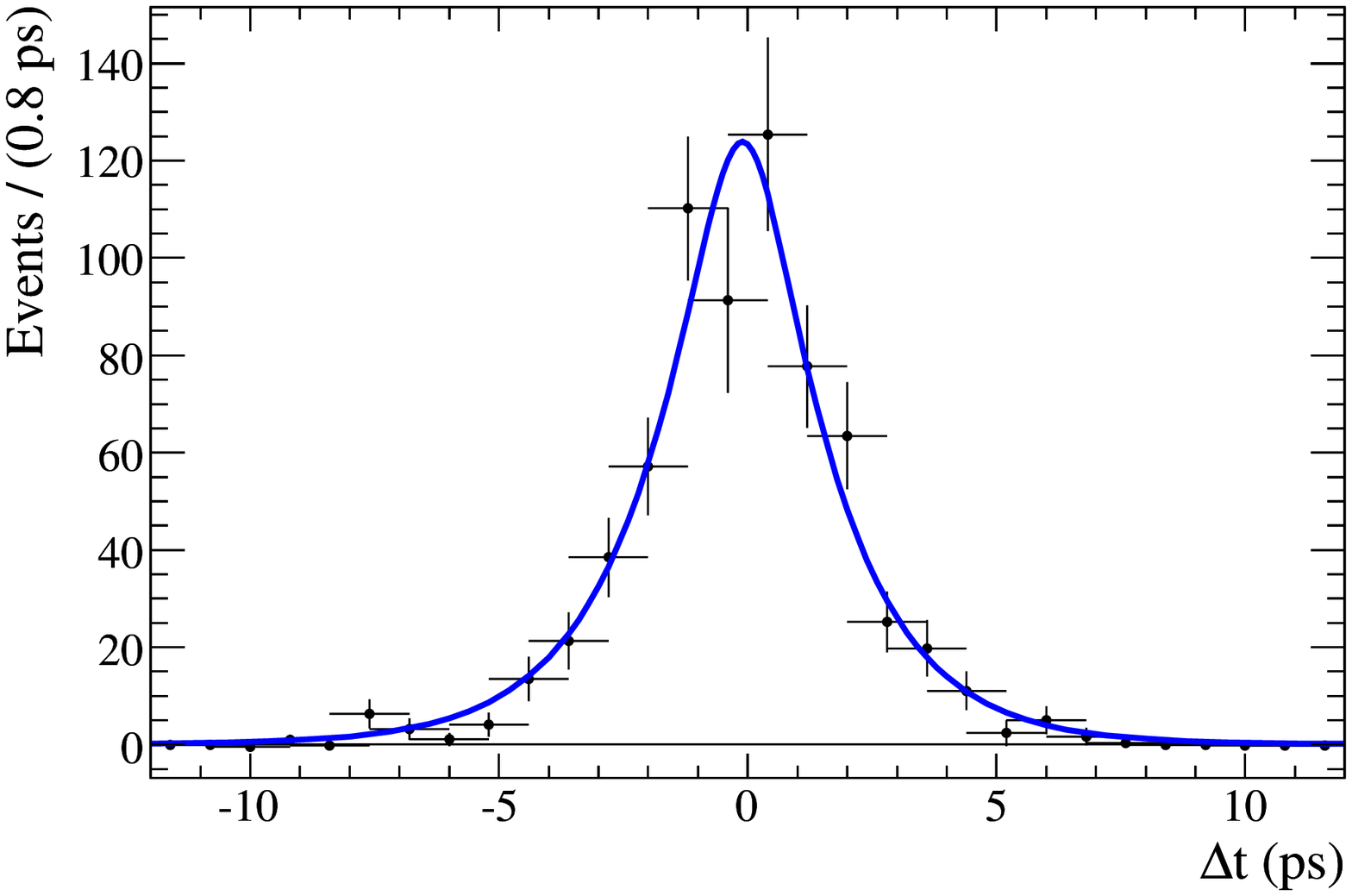}
  \includegraphics[width=0.32\linewidth]{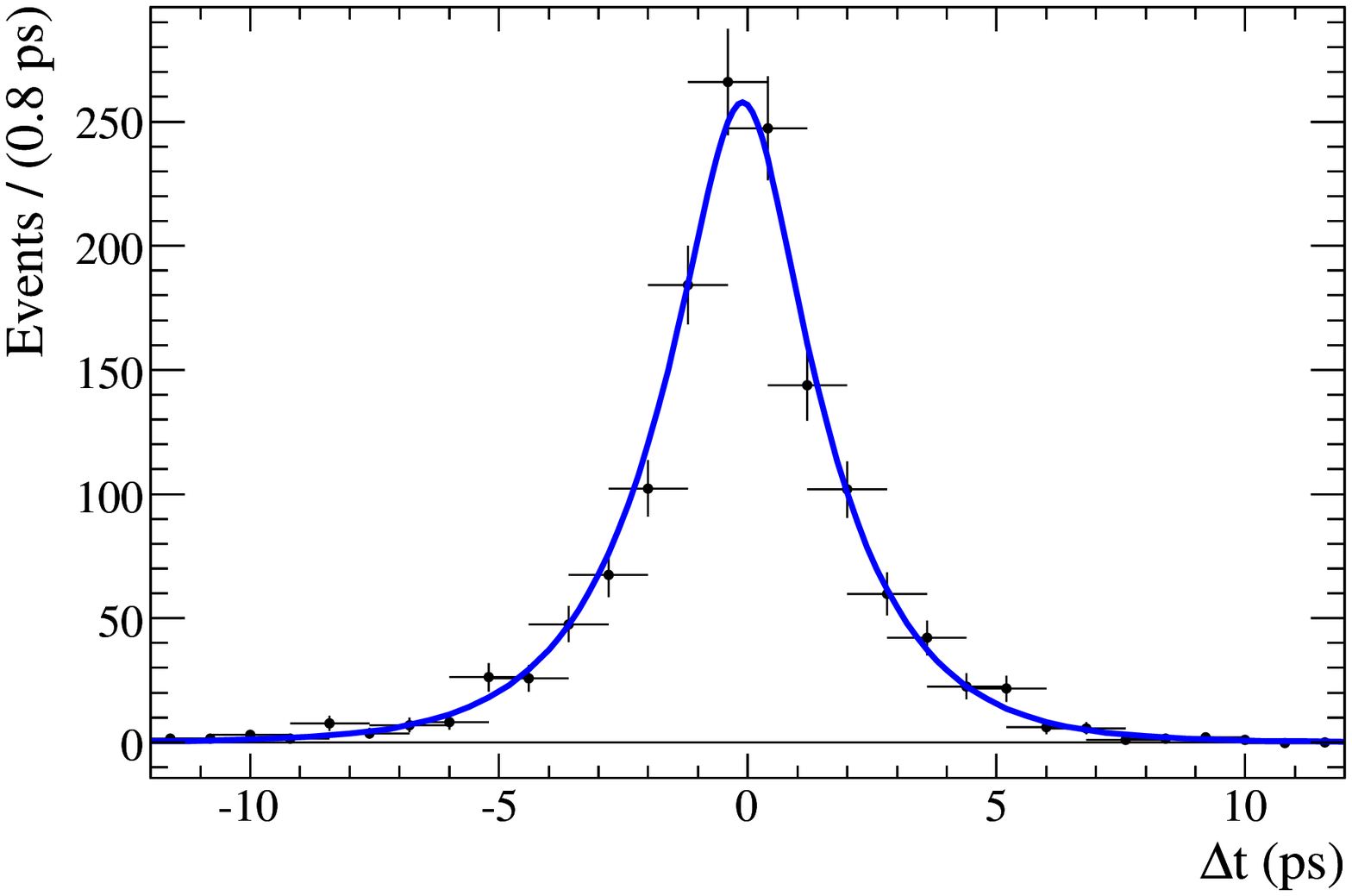}
  \includegraphics[width=0.32\linewidth]{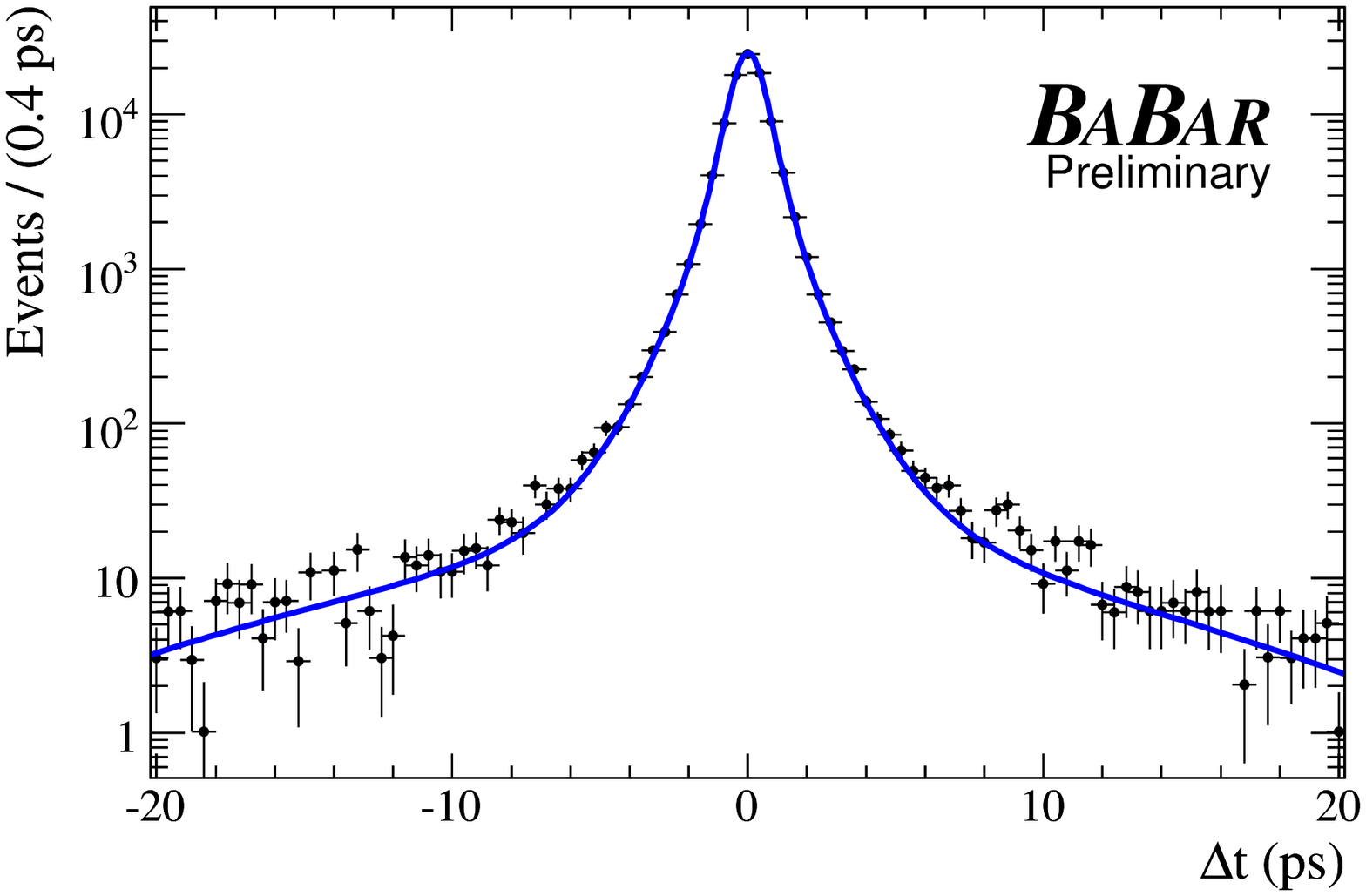}
\caption{ The  background-subtracted distributions of 
$\deltat$ for signal $\pip\pim$ (left), $\Kpm\pimp$ (middle), and 
the signal-subtracted $\deltat$ distribution for background 
candidates in the data (right).  
The curves represent the PDFs used in the fit.}
\label{fig:hhdt}
\end{center}
\end{figure}

\begin{figure}[!tbph]
\begin{center}
  \includegraphics[width=0.69\linewidth]{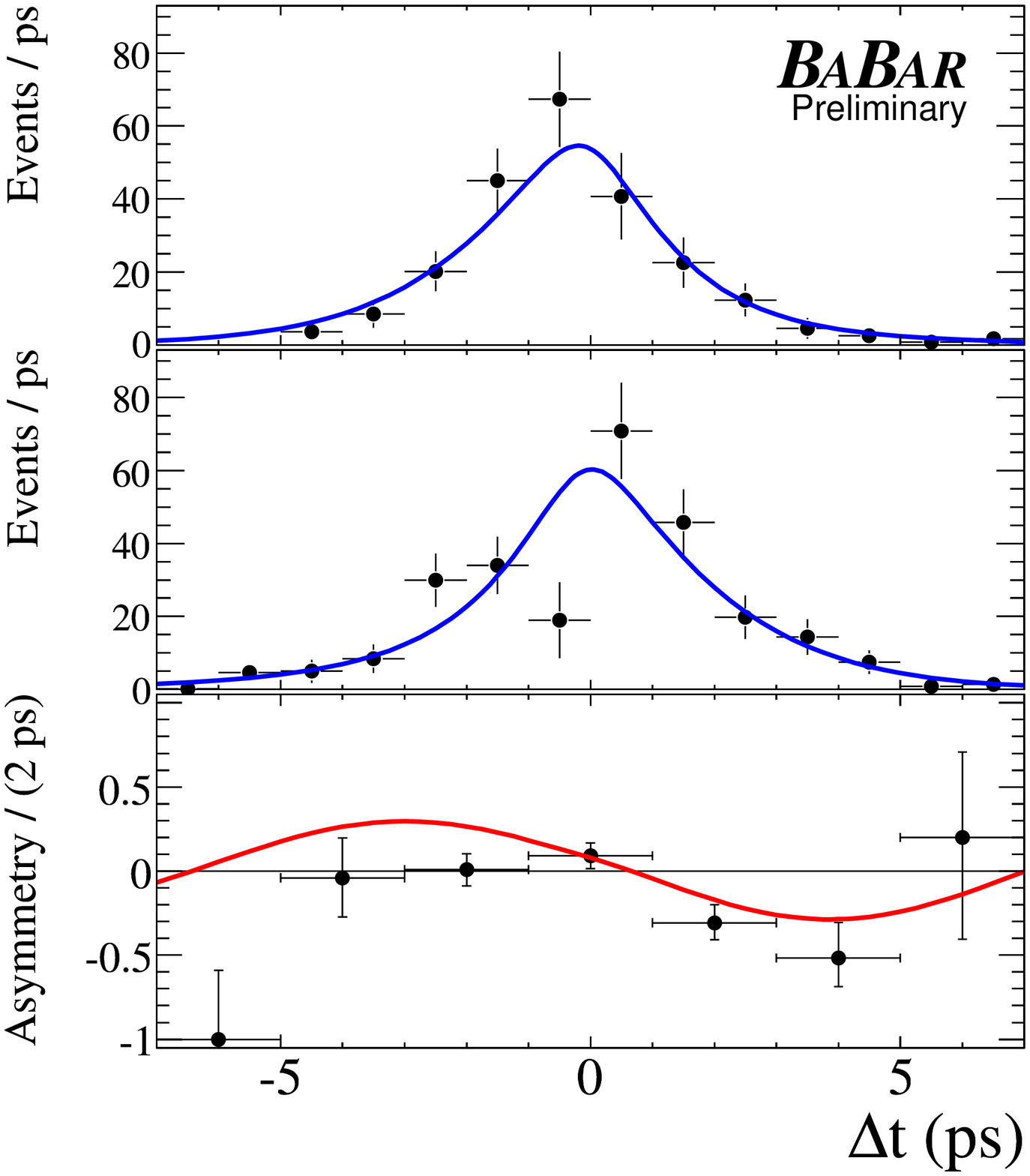}
\caption{ The background-subtracted distributions of 
$\deltat$ for signal $\pip\pim$ events tagged as $\Bz$ (top) 
or $\Bzb$ (middle), and the asymmetry, defined as 
${\cal A} = \left ( N_{\Bz} - N_{\Bzb}\right )
/\left ( N_{\Bz} + N_{\Bzb}\right )$ (bottom).
The curves represent the PDFs used in the fit.}
\label{fig:asym}
\end{center}
\end{figure}

\begin{figure}[!tbp]
\begin{center}
\includegraphics[width=0.69\linewidth]{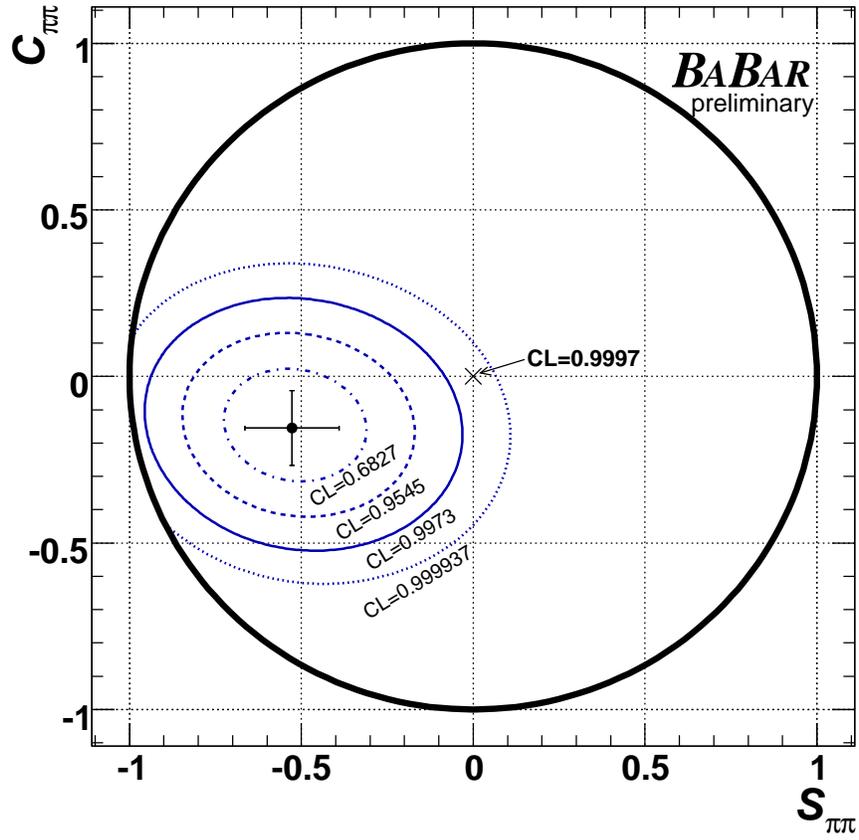}
\end{center}
\vspace{0.2cm}
\caption{ 
  Central value and errors for \spipi and \cpipi and confidence-level
  (C.L.) contours.  The measured value is $3.6~\sigma$ from the point of 
  no \CP violation ($\spipi=0$, $\cpipi=0$), converting from the C.L.\ to the units of 
  ``two-sided'' Gaussian significance.}
\label{fig:SCcontour}
\end{figure}

All results for the \Bztohh decay modes are listed in
Table~\ref{tab:resultsB}. The correlation coefficient between \spipi
and \cpipi is found to be $-0.082$.  The data distributions of $\mes$,
$\de$, and $\fish$ for \Bztohh decays are shown in
Fig.~\ref{fig:hhVarSig} for signal and Fig.~\ref{fig:hhVarBkg} for
background with the event-weighting technique. The direct \CP
asymmetry in \Bztokpi is apparent in the distribution of \de for \Bz
and \Bzb decays, shown in Fig.~\ref{fig:deakpi}.  We show the
distributions of $\deltat$ for signal and background decays in
Fig.~\ref{fig:hhdt}.  In Fig.~\ref{fig:asym}, we show the distribution
of \deltat separately for \Bztopipi events tagged as $\Bz$ or $\Bzb$,
and the asymmetry $a(\deltat)$.  The central values and errors for
\spipi and \cpipi are shown in Fig.~\ref{fig:SCcontour}, along with
confidence-level contours.  Our measurement excludes the absence of  
\CP violation ($\spipi=0, \cpipi=0$) at a confidence level of
0.99970, or $3.6~\sigma$. 
 
Systematic uncertainties for the \CP asymmetries $\akpi$, \spipi, and \cpipi
are listed in Table~\ref{tab:pipisyst}.  For the asymmetry in the
$\Kp\pim$ mode, we find a background asymmetry of $-0.0042\pm 0.0064$, which
is consistent with zero.  We therefore take the sum in quadrature of
the central value of the background asymmetry and its statistical
uncertainty as the 
systematic uncertainty 
on the signal $\akpi$ to account for possible charge-dependent
detector and analysis bias.  To further check for biases in the
fitting technique, we perform a large number of pseudo-experiments
where the signal events are randomly sampled from simulated MC events,
and the background is generated directly from the PDFs.  We find a
bias in $\akpi$ of $0.002$ and include this in the systematic
uncertainty.  The biases on $\spipi$ and $\cpipi$ in this study are
consistent with zero, so we take the sum in quadrature of the central
value and its uncertainty as the systematic error due to potential
bias in the fitter.  The remaining systematic effects for $\spipi$ and
$\cpipi$ are dominated by uncertainties in the parameterization of
$B$-flavor tagging and vertexing, and (for $\cpipi$) in the effect of
\CP\ violation on the tag side.

\section{CONCLUSIONS}
\label{sec:Conclusions}

\begin{figure}[!tbp]
\begin{center}
\includegraphics[width=0.69\linewidth]{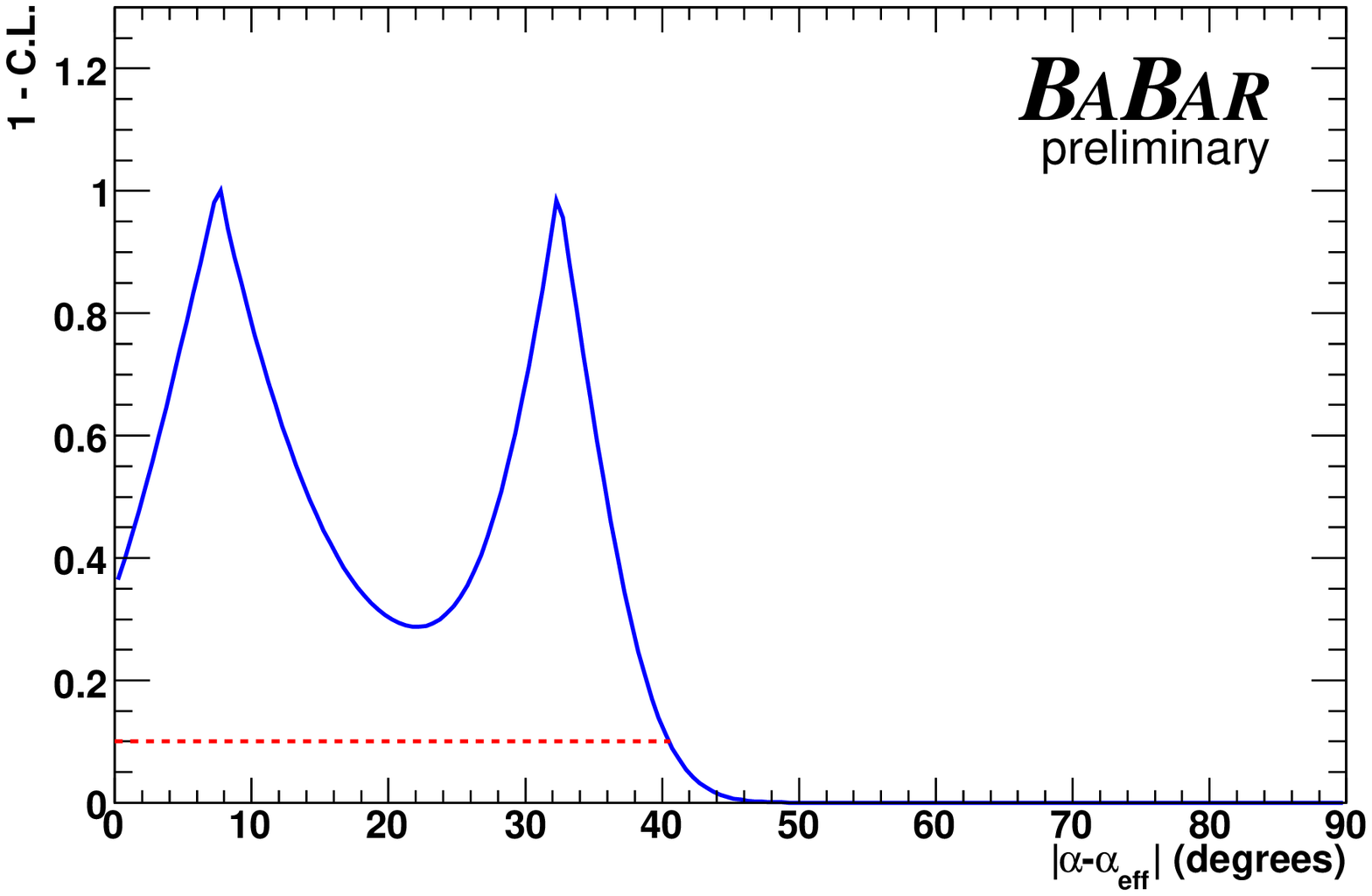}
\includegraphics[width=0.69\linewidth]{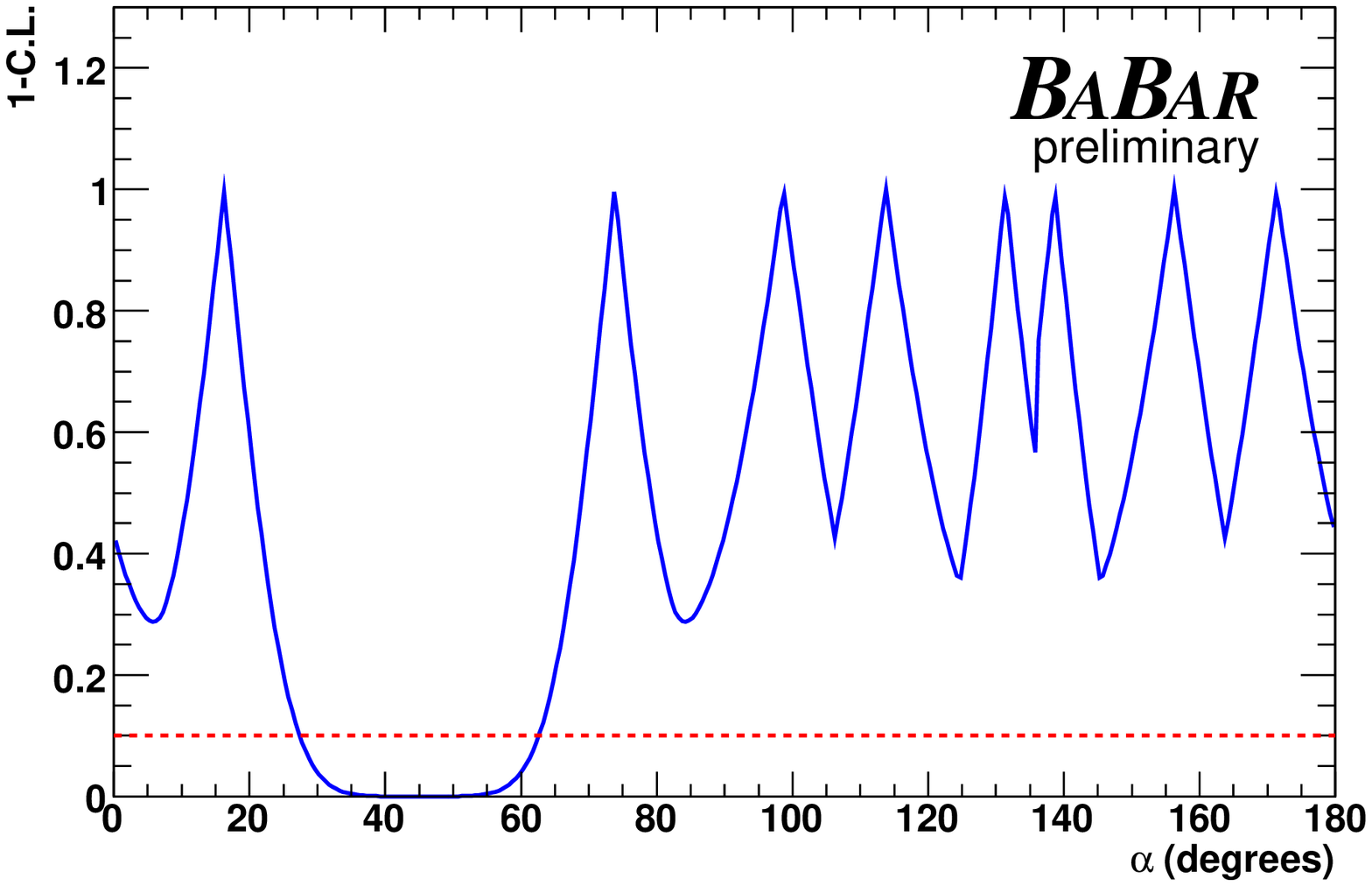}
\end{center}
\vspace{0.2cm}
\caption{
  Constraint on the angle $\delalph = \alpha-\alphaeff$ (top), expressed
  as one minus the confidence level (C.L.) as a function of $|\delalph|$.  We find an
  upper bound on $\left|\delalph\right|$ of $41^{\rm o}$ at the 90\% C.L.
  Constraint on the CKM angle $\alpha$ (bottom) expressed as $ 1 -
  {\rm C.L.}$
  The eight peaks correspond to an eight-fold ambiguity in the
  extraction of $\alpha$; four solutions are from the value and sign
  of $\delalph$, which is doubled due to the trigonometric reflections between
  $\alphaeff$ and $\pi/2 - \alphaeff$. Only the isospin-triangle
  relations and the expressions in Eqn.~\ref{eq:asymmetry} are used in
  this constraint.  The solution at exactly $\alpha=0$ is excluded 
  at $ 1 - {\rm C.L.} =  4.4\times10^{-5}$, not shown in
  the plot, corresponding
  to the exclusion of $\spipi=0$, $\cpipi=0$ at 3.6 $\sigma$. Some of the
  solutions, and the region around $\alpha = 0$, can be disfavored by
  other physics information.}
\label{fig:alpha}
\end{figure}

The branching-fraction and \CP-asymmetry results described in this
paper are:
\begin{align*}
   \spipi & =   -0.53 \pm 0.14 \pm 0.02 \\
   \cpipi & =   -0.16 \pm 0.11 \pm 0.03 \\
   {\cal A}_{K\pi} & = -0.108 \pm 0.024 \pm 0.008 \\
   \BR(\Bztopizpiz) & = ( 1.48 \pm 0.26 \pm 0.12 ) \times 10^{-6} \\
   \BR(\Btopipiz) & = ( 5.12 \pm 0.47 \pm 0.29 ) \times 10^{-6} \\
   \BR(\Btokpiz) & = ( 13.3 \pm 0.56 \pm 0.64 ) \times 10^{-6} \\
   \cpizpiz & =  -0.33 \pm 0.36 \pm 0.08   \\
   \apipiz & =  -0.019 \pm 0.088 \pm 0.014   \\
   \akpiz & =  0.016 \pm 0.041 \pm 0.012 . 
\end{align*}
Combining these with a branching fraction 
$\BR(\Bztopippim) = (5.8 \pm 0.4 \pm 0.3) \times 10^{-6}$, 
also measured by \babar~\cite{BabarBRPiPi}, we may evaluate the
constraints on both the penguin contribution to $\alpha$ and on the
CKM angle $\alpha$ itself. Constraints are evaluated by scanning the
parameters of interest, $\left|\delalph\right| = |\alpha - \alphaeff|$ and $\alpha$, and
then calculating the $\chi^{2}$ for the five amplitudes ($A^{+0}$,
$A^{+-}$, $A^{00}$, $\tilde{A}^{+-}$, $\tilde{A}^{00}$) given our
measurements and the isospin-triangle relations~\cite{ref:CKMfitter}. The $\chi^2$ is
converted to a confidence level (C.L.) as shown in Fig.~\ref{fig:alpha}.  The
upper bound on  $\left|\delalph\right|$ is $41^{\rm o}$ at the 90\%
C.L. Somewhat more restrictive new constraints on $\alpha$ are found from
measurements of $B\to\rho\rho$ and $B\to\rho\pi$ decays~\cite{ref:rhorho}.

We have also presented updated preliminary measurements of the branching
fraction for $\Kp\piz$, and the charge asymmetries in $\Kp\pim$ and $\Kp\piz$.
Ignoring color-suppressed tree amplitudes, the charge asymmetries in $\Kp\pim$ 
and $\Kp\piz$ should be equal (see Gronau and Rosner in Ref.~\cite{ref:SumRule}), which has 
not been supported by recent data~\cite{ref:BaBarKpiData}.  
The values of $\akpi$ and ${\cal A}_{K\piz}$ reported here are 
separated by over two standard deviations.  These results could 
indicate a large color-suppressed amplitude, an enhanced electroweak penguin,
or possibly new-physics effects~\cite{ref:NP}.

\section{ACKNOWLEDGMENTS}
\label{sec:Acknowledgments}

We are grateful for the 
extraordinary contributions of our \pep2\ colleagues in
achieving the excellent luminosity and machine conditions
that have made this work possible.
The success of this project also relies critically on the 
expertise and dedication of the computing organizations that 
support \babar.
The collaborating institutions wish to thank 
SLAC for its support and the kind hospitality extended to them. 
This work is supported by the
US Department of Energy
and National Science Foundation, the
Natural Sciences and Engineering Research Council (Canada),
Institute of High Energy Physics (China), the
Commissariat \`a l'Energie Atomique and
Institut National de Physique Nucl\'eaire et de Physique des Particules
(France), the
Bundesministerium f\"ur Bildung und Forschung and
Deutsche Forschungsgemeinschaft
(Germany), the
Istituto Nazionale di Fisica Nucleare (Italy),
the Foundation for Fundamental Research on Matter (The Netherlands),
the Research Council of Norway, the
Ministry of Science and Technology of the Russian Federation, 
Ministerio de Educaci\'on y Ciencia (Spain), and the
Particle Physics and Astronomy Research Council (United Kingdom). 
Individuals have received support from 
the Marie-Curie IEF program (European Union) and
the A. P. Sloan Foundation.

\clearpage

\begin{table}[!phtb]
\caption{Average tagging efficiency $\epsilon$, average mistag fraction $w$,
mistag fraction difference $\Delta w = w(\Bz) - w(\Bzb)$, and effective tagging efficiency 
$Q$ for signal events in each tagging category.  The quantities are measured in the 
\Bflav\ sample.}
\smallskip
\begin{center}
\begin{tabular}{crclrclrclrcl} \hline\hline
Category & \multicolumn{3}{c}{$\epsilon\,(\%)$} & \multicolumn{3}{c}{$w\,(\%)$} & \multicolumn{3}{c}{$\Delta w\,(\%)$} &
\multicolumn{3}{c}{$Q\,(\%)$} \rule[-2mm]{0mm}{6mm} \\\hline
{\tt Lepton}    & $8.67 $&$ \pm $&$ 0.08  $&$ 3.0  $&$ \pm $&$ 0.3 $&$ -0.2 $&$ \pm $&$ 0.6 $&$ 7.7  $&$ \pm $&$ 0.1$\\
{\tt Kaon\,I}   & $11.0 $&$ \pm $&$ 0.08  $&$ 5.3  $&$ \pm $&$ 0.4 $&$ -0.6 $&$ \pm $&$ 0.7 $&$ 8.7  $&$ \pm $&$ 0.2$\\
{\tt Kaon\,II}  & $17.2 $&$ \pm $&$ 0.1   $&$ 15.5 $&$ \pm $&$ 0.4 $&$ -0.4 $&$ \pm $&$ 0.7 $&$ 8.2  $&$ \pm $&$ 0.2$\\
{\tt Kaon\,Pion}& $13.8 $&$ \pm $&$ 0.09  $&$ 23.5 $&$ \pm $&$ 0.5 $&$ -2.4 $&$ \pm $&$ 0.8 $&$ 3.9  $&$ \pm $&$ 0.1$\\
{\tt Pion}      & $14.4 $&$ \pm $&$ 0.09  $&$ 33.0 $&$ \pm $&$ 0.5 $&$  5.2 $&$ \pm $&$ 0.8 $&$ 1.7  $&$ \pm $&$ 0.1$\\
{\tt Inclusive} & $9.6  $&$ \pm $&$ 0.08  $&$ 41.9 $&$ \pm $&$ 0.6 $&$  4.6 $&$ \pm $&$ 0.9 $&$ 0.25  $&$ \pm $&$ 0.04$\\
{\tt Untagged}  & $ 23.4 $&$ \pm $&$ 0.12 $\\\hline
Total $Q$       &        &       &       &        &       &       &        &       &       &$ 30.4 $&$ \pm $&$ 0.3$ \rule[-2mm]{0mm}{6mm} \\\hline\hline
\end{tabular}
\end{center}
\label{tab:tagging}
\end{table}

\begin{table*}[!tbh]
\caption{ 
  The results for the \Bztopizpiz and \Btohpiz decay modes are summarized.  For each
  mode, the number of signal events $N_{\rm S}$, total detection efficiency
  $\varepsilon$, branching fraction \BR, and \CP asymmetry are given.
  Errors on $\BR$ and the asymmetries are
  statistical and systematic, respectively, while errors for $N_{\mathrm{S}}$ are statistical and those 
   for $\varepsilon$ are purely systematic.}
\label{tab:resultsA}
\smallskip
\begin{center}
\begin{tabular}{l|ccccc} \hline \hline
Mode        &  $N_{\mathrm{S}}$      & $\varepsilon$ (\%) & \BR($10^{-6}$)        & Asymmetry                               \\ \hline 
\Bztopizpiz &  $140\pm 25$  & $27.1\pm 1.7$      &  $1.48\pm 0.26\pm 0.12$ &  $\cpizpiz = -0.33  \pm 0.36 \pm 0.08$   \\
\Btopipiz   &  $572\pm 53$  & $32.1\pm 1.8$       &  $5.12\pm 0.47\pm 0.29$ &  $\apipiz  = -0.019 \pm 0.088 \pm 0.014$ \\
\Btokpiz    &  $1239\pm 52$ & $26.8\pm 1.3$       &  $13.3\pm 0.56\pm 0.64$ &  $\akpiz   =  0.016 \pm 0.041 \pm 0.012$  \\
\hline\hline
\end{tabular}
\end{center}
\end{table*}

\begin{table*}[!hptb]
\caption{ 
  Systematic uncertainties in the determination of the \Bztopizpiz
  branching fraction (left) as a percentage change, and the $C_{\piz\piz}$ asymmetry
  (right) as an absolute change.}
\label{tab:pizpizsyst}
\smallskip
\begin{center}
\begin{tabular}{l|cc}
Source  &  $\Delta \BR (\piz\piz)$   \\ \hline \hline
\piz efficiency        & $ 6.0\%$ \\
\de resolution         & $ 3.6\%$ \\
\mes PDF endpoint      & $ 3.1\%$ \\
mean of \de and \mes   & $ 1.9\%$  \\
\BR (\Btorhopi)        & $ 1.8\%$  \\
luminosity             & $ 1.1\%$  \\ \hline
Total                  & $ 8.2\%$ \\ \hline\hline
\end{tabular}
\hspace*{1.25cm}
\begin{tabular}{l|c}
Source  &  $\Delta (C_{\piz\piz})$         \\ \hline \hline
tagging                 &       $ 0.06$ \\
\mes and \de            &       $ 0.04$ \\
\B background asymmetry &       $ 0.03$ \\ \hline
Total                   & $ 0.08$ \\ \hline\hline
\end{tabular}
\end{center}
\end{table*}

\begin{table}[!tbhp]
\caption{ Dominant systematic uncertainties for \Btopipiz and
  \Btokpiz, as percentage changes in the branching
  fractions \BR \ (left), and absolute changes in the asymmetries
  $\acp_{\pipm\piz}$, $\acp_{K^\pm\piz}$ (right).  ${\cal A} = \left ( N_{\Bz} - N_{\Bzb}\right )
/\left ( N_{\Bz} + N_{\Bzb}\right )$.}
\label{tab:hpizsyst}
\smallskip
\begin{center}
\begin{tabular}{l|cccc}
Source                  & $\Delta \BR (\pi^\pm\piz)$ & $\Delta \BR (K^\pm\piz)$ \\ \hline \hline
\mes and \de res. &    $ 3.1  \%$           &    $ 2.4\%$  \\
\fish PDF               &    $ 3.1  \%$           &    $ 2.1\%$  \\
\piz efficiency         &    $ 3.0  \%$           &    $ 3.0\%$  \\
\de mean                &    $ 1.2  \%$           &    $ 3.0\%$  \\
fit bias                &    $ 1.1  \%$           &    $ 1.8\%$  \\
\B background           &    $ 0.7  \%$           &    $ 0.2\%$  \\
$h^\pm$  identification &    $ 0.7  \%$           &    $ 0.8\%$  \\ \hline
Total                   &    $ 5.6  \%$           &    $ 4.8\%$  \\ \hline \hline
\end{tabular}
\hfill
\begin{tabular}{l|cc}
Source                  &   $\Delta(\apipiz)$  &  $\Delta(\akpiz)$ \\ \hline \hline
\mes and \de            &   $ 0.007$        &  $ 0.003$ \\
\B background           &   $ 0.009$        &  $ 0.002$ \\
detector asymmetry      &   $ 0.008$        &  $ 0.011$ \\ \hline
Total                   &   $ 0.014$        &  $ 0.012$   \\ \hline \hline
\end{tabular}
\end{center}
\end{table}

\begin{table*}[!tbhp]
\caption{ 
  The results for the \Bztohh decay modes are summarized.  For each
  mode, the number of signal events $N_{\rm S}$ and \CP asymmetries are shown.  
  Errors are statistical and systematic.}
\label{tab:resultsB}
\smallskip
\begin{center}
\begin{tabular}{l|cc}
\hline\hline
Mode        & $N_{\rm S}$      & Asymmetry       \\ \hline 
\Bztopippim &  $675\pm 42$& $\spipi = -0.53 \pm 0.14 \pm 0.02;  \;\; \cpipi = -0.16 \pm 0.11 \pm 0.03$ \\
\Bztokpi    &  $2542\pm 67$ & $\akpi  = -0.108 \pm 0.024 \pm 0.008$ \\
\Bztokk     &  $11\pm 19$ & ---  \\
\hline\hline
\end{tabular}
\end{center}
\end{table*}

\begin{table}[!bthp]
\caption{
Summary of systematic uncertainties on $\akpi$, $\spipi$, and $\cpipi$.  
The total uncertainty is calculated as the sum in quadrature of the 
individual contributions.}
\label{tab:pipisyst}
\begin{center}
\begin{tabular}{lccc} \hline\hline
Source                      & $\akpi$ & $\spipi$ & $\cpipi$ \\
\hline
PDF parameters              & $ 0.0007$ & $0.0003$    & $0.0011$ \\
Tagging/Vertexing           & ---       & $0.0178$    & $0.0170$ \\
SVT alignment               & ---       & $0.0100$    & $0.0022$ \\
Beam spot                   & ---       & $0.0100$    & $0.0100$ \\
Tag-side interference       & ---       & $0.0080$    & $0.0230$ \\
$\tau_{\Bz}$ and $\deltamd$ & ---       & $0.0015$    & $0.0036$ \\
Potential bias              & $0.0078$  & $0.0051$    & $0.0036$ \\
\hline
Total                       & $ 0.0078$ & $ 0.0247$   & $0.0300$ \\
\hline\hline
\end{tabular}
\end{center}
\end{table}

\end{document}